\documentclass[%
aip,
amsmath,amssymb,
reprint,%
]{revtex4-2}
\usepackage{graphicx}
\usepackage{dcolumn}
\usepackage{mathrsfs}
\usepackage{bm}
\usepackage{lineno}
\usepackage{hyperref}
\usepackage{bbm}
\usepackage{ulem}
\usepackage{xcolor}

\begin{document}
\title{Optical interference by amplitude measurement }

\author{Yunxiao Zhang*}
\affiliation{State Key Laboratory of Precision Measurement Technology and Instruments, College of Precision Instrument and Opto-Electronics Engineering, Tianjin University, Tianjin 300072, P. R. China}

\author{Xuan Tang*}
\affiliation{Department of Physics, City University of Hong Kong, 83 Tat Chee Avenue, Kowloon, Hong Kong, P. R. China}

\author{Xueshi Guo}
\affiliation{State Key Laboratory of Precision Measurement Technology and Instruments, College of Precision Instrument and Opto-Electronics Engineering, Tianjin University, Tianjin 300072, P. R. China}

\author{Liang Cui}
\affiliation{State Key Laboratory of Precision Measurement Technology and Instruments, College of Precision Instrument and Opto-Electronics Engineering, Tianjin University, Tianjin 300072, P. R. China}

\author{Xiaoying Li}
 \email{xiaoyingli@tju.edu.cn}
\affiliation{State Key Laboratory of Precision Measurement Technology and Instruments, College of Precision Instrument and Opto-Electronics Engineering, Tianjin University, Tianjin 300072, P. R. China}

\author{Z. Y. Ou}
 \email{jeffou@cityu.edu.hk}
\affiliation{Department of Physics, City University of Hong Kong, 83 Tat Chee Avenue, Kowloon, Hong Kong, P. R. China}

\begin{abstract}
* equal contributions\\

\noindent Interference effects are usually observed by intensity measurement. Path indistinguishability by quantum complementarity principle requires projection of the interfering fields into a common indistinguishable mode before detection. On the other hand, the essence of wave interference is the addition of amplitudes of the interfering fields. Therefore, if amplitudes can be directly measured and added, interference can occur even though the interfering fields are in well-distinguishable modes. Here, we make a comprehensive study in both theory and experiment of a technique by homodyne measurement of field amplitudes to reveal interference. This works for both classical and quantum fields even though there exists distinguishability in the interfering paths of light. This directly challenges complementarity principle. We present a resolution of this issue from the viewpoint of measurement that emphasizes either particle or wave. This technique is particularly useful for recovering interference in unbalanced interferometers with path-imbalance beyond coherence length of the input field and can be applied to remote sensing to extend applicable range.  Since the amplitude-based interference phenomena studied here are fundamentally different from the traditional intenisty-based interference phenomena, our approach leads to a new paradigm to study coherence between optical fields. 

\end{abstract}

\maketitle

\section{Introduction}

Light, as an optical wave like all other waves, exhibits interference effects due to wave superposition.
At the fundamental level, the conditions for observing optical interference are summarized as the quantum complementarity principle: indistinguishability in optical fields gives rise to interference \cite{bohr,dis}. Indistinguishability, on the other hand, is characterized by the modes of optical fields such that fields in orthogonal modes will have complete distinguishability and exhibit no interference between them. An example is the superposition of fields in orthogonal x and y polarization. Another example is pulsed fields in non-overlapping temporal modes: when pulses of two fields are well separated (temporal modes are orthogonal), no interference can be observed because of temporal distinguishability. This applies to cw fields as well in a path-unbalanced interferometer beyond the coherence length of the fields. For quantum fields, distinguishability can also be obtained through entanglement \cite{zou91,zei94}.

The requirement of indistinguishability is due to the fact that detection of optical fields is intensity measurement which is the square modulus of amplitude. So, there must be amplitude addition of the fields before detection to preserve the phase information of the fields. The common way to recover interference is the projection of the interfering fields on to a common mode before detection, either directly or indirectly via projection measurement \cite{zei94}. In the example of superposition of fields with orthogonal x and y polarization, a polarizer at 45 degree in front of the detector can restore interference. In the case of well-separated pulses, a narrow band optical filter can lengthen the pulse width so that they will overlap after the filter to give rise to interference.  For cw fields, spectral filtering can lengthen the coherence length so that interference can occur in frequency domain in the path unbalanced interferometer \cite{wolf,agar,mandel93}. Such a technique even applies to matter waves in neutron interferometers \cite{rauch}. 

The aforementioned methods to recover interference still require physical addition of optical waves in a common mode and the added waves must be coherent to each other, which requires balance of interference path to within the coherence length of the fields before the intensity measurement, still leading to the indistinguishability requirement to reveal interference. It should be noted that the projection to common modes must be done before detection since phase information is lost after intensity measurement.  In fact, the traditional optical coherence theory, in both classical wave \cite{mw} and quantum \cite{gl} form, is based on measurement of the field intensity as the obseravble quantity because the fast oscillation (frequency $\sim 10^{15}$ Hz) of the electromagnetic wave in optical regime makes it impossible to directly measure the field quantities like electric or magnetic field \cite{wolf54}.
 Intensity measurement relies on photo-electric effect, which shows the particle aspect of light or in other words, makes projection of quantum system onto photon number states, and totally loses the wave information of light. It is well-known that observation of interference highly depends how we make measurement. Therefore, the requirement of projection to common modes in intensity-based interference can be viewed as  a result of the measurement emphasizing on the particle aspect of light. 

So, is there a measurement process that emphasizes the wave aspect of light field? As is well-known, amplitude is a characteristics of wave and if it can be directly measured, such as in the case of radio frequency electromagnetic (EM) waves where radio antenna directly picks up EM wave signals and converts to electric currents, it is possible to achieve wave interference in the addition of the electric currents without the need to bring the wave physically together. This is the working principle of the very long baseline array of radio telescopes for high resolution imaging at RF band \cite{syn}. Unfortunately, because of the high frequency in optical regime, optical antenna does not exists. On the other hand, optical homodyne detection technique by mixing incoming field with a strong coherent local oscillator field is a measurement process that measures the amplitude of the optical wave \cite{sh} and it was shown that its output photo-current corresponds exactly to the quantum measurement of the operator of the quadrature-phase amplitude of an optical field \cite{ou-kim}. Since superposition of waves is the addition of their amplitudes and thus interference is in essence the coherent addition of the wave amplitudes, the addition of the photo-currents from homodyne detection will give the direct addition of the amplitudes of the fields and should lead to interference.   Indeed, this was first demonstrated recently in the recovery of interference in an unbalanced SU(1,1) interferometer by homodyne detection\cite{huo22}.

Historically, the technique of optical homodyne detection (HD) has been extensively studied since 1980s. This technique brings the tools of grafting radio technology into optical information processing, as is evidenced by the rapid progress made in the field of coherent optical communication in the past two decades. The advantage is in its higher capacity of information by encoding and decoding in both amplitude and phase \cite{tay,gfli}. At the current stage,  this application of HD requires a steady phase relation between the local oscillator and the field to be measured, which is also true in other similar applications such as optical coherence tomography and LIDAR \cite{oct}. This adds the complexity to the technique of homodyne detection. On the other hand, with the advent of quantum optics in 1980's, homodyne detection is best known to measure the quantum noise of light field  to reveal quantum nature of light \cite{sh,ou-kim}. In analogy to radio astronomy, homodyne/heterodyne detection technique was also used in infrared stellar interferometry for high resolution astronomy \cite{joh,isi}. The technique was not pursued further because the shot noise/vacuum noise background in optical regime overwhelms the weak celestial light signal.
Therefore, the intensity measurement, because of its simplicity, is mostly the measurement technique in optical interferometry, and the aforementioned projection to common mode is the only way for amplitude addition in optical fields. This then leads to the indistinguishability requirement that has to be met in order to see interference in optical regime. This is perhaps why optical interference is not a commonly seen phenomenon in our daily life even though light is everywhere, in contrast to other waves such as sound waves and water waves. 

In this paper, we consider the homodyne detection technique in general for the observation of interference of various fields. Because we add photo-currents for direct amplitude addition, we can relax the requirement discussed earlier for intensity measurement and there is no need to project the two interfering fields onto the same mode to reveal interference effect. Therefore, two orthogonal fields, such as x and y polarized fields or temporally non-overlapping fields in path-unbalanced interferometer, can exhibit interference as long as they have phase correlation.  Different from the technique used in infrared stellar interferometry mentioned above, our approach uses only one homodyne detection but achieve amplitude addition of two fields for interference.

Although homodyne detection technique is not suitable for studying photon correlation due to large contribution of vacuum noise at low photon number, quantum states of light are known to suppress vacuum noise and lead to enhanced signal-to-noise in optical interferometric applications for optical sensing \cite{xiao87,slu87}. So, applying homodyne detection technique to studying quantum interference with the employment of quantum states is a natural choice of experimental tools. Although the technique emphasizes the wave nature because amplitudes are associated only with wave, it applies equally well to single-photon state, and thus reveals further the particle-wave duality of light.

Another surprising consequence of  observing interference based on amplitude measurement is that the traditional concepts about coherence may need to revise. For example, as is well known, interferometers require balance of the two interfering paths to within the coherence length of the input field, which is the limiting factor for Michelson's stellar interferomtry \cite{mich}. This is based on superposition of optical waves before intensity measurement. For the new technique of amplitude-based interference, no requirement is needed for the superposition of fields at one location such as the beam splitter of a Mach-Zehnder interferometer, so it is particularly useful for unbalanced interferometers beyond coherence length, which have recently gathered some attentions due to potential applications in remote sensing \cite{njp,kim,ou22}. In these studies, the higher-order correlation measurement technique of coincidence measurement is used. But here we do not use photon coincidence measurement at two locations but homodyne detection at one location.

The paper is organized as follows. For the general readers to understand the basic idea of observing interference by homodyne detection, we first use classical wave theory in Sec. II to discuss a simple scheme of interference of fields with orthogonal  polarizations. We then extend the idea to unbalanced interferometer beyond coherence length in Sec. III and to pulsed cases in Sec.IV. In each case, we provide experimental demonstration in support of the theoretical ideas. In Section V, we apply to quantum fields to demonstrate the recovery of interference and discuss the effect of quantum noise.   In Section VI, we discuss the connection of homodyne detection to the fourth-order interference effects and conclude with a discussion in Sec.VII.

\section{General Principle and its experimental verification}

In order to demonstrate the general principle behind the new  amplitude-based interference scheme in a simple way, we first discuss a scheme with classical fields in orthgogonal polarization states and extend in later sections to its more complicated variations as well as to quantum states of light by using quantum theory.
\vskip 0.3 in 
\noindent {\bf Theory}

\begin{figure}[t]
\includegraphics[width=6.5cm]{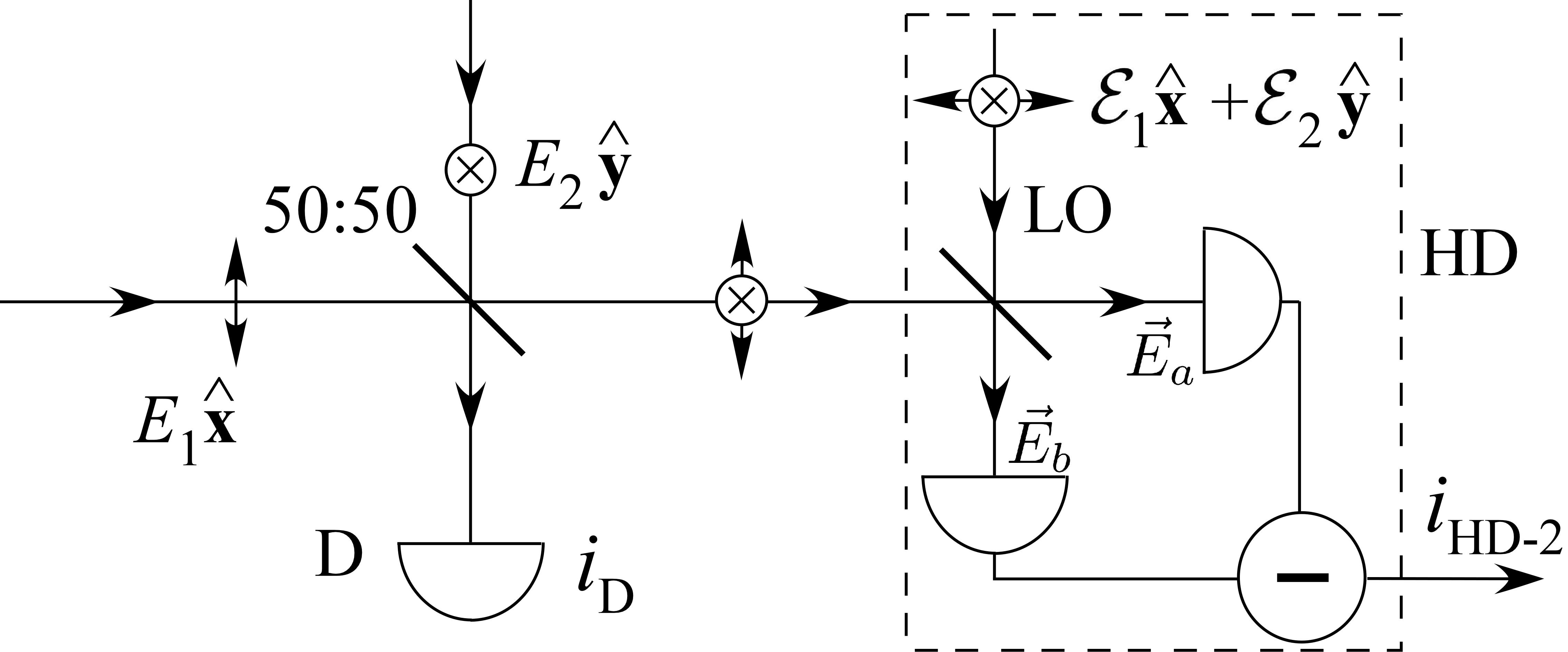}
	\caption{Interference between two orthogonally polarized fields. HD: homodyne detection.}
	\label{XY}
\end{figure}

We start by considering superposition of two fields in orthogonal modes. For simplicity without loss of generality, we set the two orthogonal modes as $\bf x$ and $\bf y$-polarization modes, $E_1\hat {\bf x} , E_2\hat {\bf y}$, and superimpose them with a 50:50 beam splitter as shown in Fig.\ref{XY}:
\begin{eqnarray}\label{pol}
\vec E = (E_1\hat {\bf x} + E_2\hat {\bf y})/\sqrt{2}.
\end{eqnarray}
Note that we ignore the spatial and temporal parts of the fields because we only concern polarization here. Obviously, there is no interference by direct intensity measurement with detector D due to orthogonality:
\begin{eqnarray}\label{int}
i_{D}\propto I = |\vec E |^2&=& \frac{1}{2} (E_1^*\hat {\bf  x} + E_2^*\hat {\bf y})\cdot (E_1\hat {\bf x} + E_2\hat {\bf y})\cr & =& (|E_1|^2+|E_2|^2)/2.
\end{eqnarray}

We next consider direct amplitude addition by using homodyne detection (HD) for the measurement of quadrature-phase amplitude of an optical field. It is well-known that the output photo-current of balanced HD has the form of  
\begin{eqnarray}\label{HD}
i_{HD} \propto  |{\cal E}| X(\varphi)
\end{eqnarray}
with ${\cal E} \equiv |{\cal E}|e^{i\varphi}$ as the amplitude of the local oscillator and $X(\varphi) = Ee^{-i\varphi}+E^*e^{i\varphi}$ as the quadrature-phase amplitude of field $E$. So, the output of HD gives directly the quadrature-phase amplitude of the incoming field.  Notice that homodyne detection does not measure directly the complex amplidue of $E(t)$ but the real quantity $X(\varphi)$. 

However, an important property of homodyne detection that is different from traditional intensity measurement is the mode selectivity by local oscillator (LO). It only measures the amplitude of the field with the same mode as the local oscillator.  Since it relies on the interference with the LO field, the outcome looses the vector nature of the input field. To have contributions from both polarization modes $E_1\hat {\bf x}, E_2\hat {\bf y}$, we need two local oscillators in ${\bf x}$ and ${\bf y}$-polarization modes:
\begin{eqnarray}\label{LO}
\vec {\cal E} = {\cal E}_1\hat {\bf x} + {\cal E}_2\hat {\bf y}
\end{eqnarray}
where ${\cal E}_j = |{\cal E}_j|e^{i\varphi_j} (j=1,2)$. For balanced homodyne detection (BHD), the fields in front of the detectors are
\begin{eqnarray}\label{Eab}
{\vec E}_a =(\vec {\cal E}+ \vec E)/\sqrt{2}, 
{\vec E}_b = (\vec {\cal E}- \vec E)/\sqrt{2},
\end{eqnarray}
and the output photo-current from BHD is then
\begin{eqnarray}\label{HD2}
i_{\rm HD-2} &\propto & {\vec E}_a ^{*}\cdot {\vec E}_a - {\vec E}_b ^{*}\cdot {\vec E}_b\cr
&=& |{\cal E}_1| X_1(\varphi_1)+ |{\cal E}_2| X_2(\varphi_2)\cr
& = & |{\cal E}_1|\big[X_1(\varphi_1)+ \sqrt{\lambda} X_2(\varphi_2)\big]
\end{eqnarray}
with $\lambda \equiv |{\cal E}_2/{\cal E}_1|^2$.  Notice that even though the input fields are orthogonal in polarization, the output current still gives an addition of the amplitudes of the two orthogonal fields. We generalize the above to the case of more than two orthogonal modes. With mode-matched LOs, the output photo-current is the sum of the amplitudes of all modes. The detail of the derivation is given in Appendix A.

With amplitude addition in the photo-current, we should be able to see interference. But normally the interfering fields are independent of the LO fields so that $\langle X \rangle = 0$ or the average current is zero.  On the other hand, in homodyne detection, we usually measure  the  power of the photo-current, which is proportional to the average of $i_{HD-2}^2$:
\begin{eqnarray}\label{HD2-i}
\langle i_{\rm HD-2}^2\rangle &\propto & \langle[X_1(\varphi_1)+ \sqrt{\lambda} X_2(\varphi_2)]^2\rangle \cr
&=& 2 [I_1+\lambda I_2+\sqrt{\lambda I_1I_2}(\gamma_{12} e^{i\Delta\varphi}+\gamma_{12}^*e^{i\Delta\varphi})] \cr
 &=& 2 (I_1+\lambda I_2)[1+ {\cal V} \cos(\Delta\varphi+\phi_{\gamma})]
\end{eqnarray}
where  $I_j\equiv \langle |E_j|^2\rangle (j=1,2), \gamma_{12}\equiv \langle E_1E_2^*\rangle/\sqrt{I_1I_2}\equiv |\gamma_{12}|e^{i\phi_{\gamma}}$,  $\phi_{\gamma} = {\rm Arg}(\gamma_{12})$, ${\cal V}\equiv  2|\gamma_{12}|\sqrt{\lambda I_1I_2}/(I_1+\lambda I_2)$, and $\Delta\varphi\equiv \varphi_1-\varphi_2$, which is the phase difference between the two LO fields. In the derivation here, we assume each field has a random phase relative to its corresponding LO field so that $\langle X_j(\varphi_j)\rangle = 0 (j=1,2)$ but there is a phase correlation between the two fields so that $\gamma_{12}\ne 0$. Notice that the interference fringe shown in Eq.(\ref{HD2-i}) depends on the phase difference $\Delta \varphi$ between two LOs, indicating the involvement of the LOs, while the phase difference between the two interfering fields appears in $\phi_{\gamma}$. A steady fringe does not require a fixed phase relation between the input field and the LO fields.

The result in Eq.(\ref{HD2-i}) clearly shows interference and the visibility is directly related to the coherence function $\gamma_{12}$ between the two orthogonal fields as if we had direct interference between the two fields even though they have orthogonal polarizations, and no coherence is required between LOs and the interfering fields. Another observation from the visibility is that it also depends the quantity $\lambda$ which is the ratio between the intensities of the two local oscillators, pointing again to the involvement of the LOs. Notice the disappearance of cross terms between $x$ and $y$ components of the fields due to orthogonality of polarizations, which also rules out interference between two LOs.
\vskip 0.1in 
\noindent {\bf Experimental verification}

\begin{figure}[t]
\includegraphics[width=7.5cm]{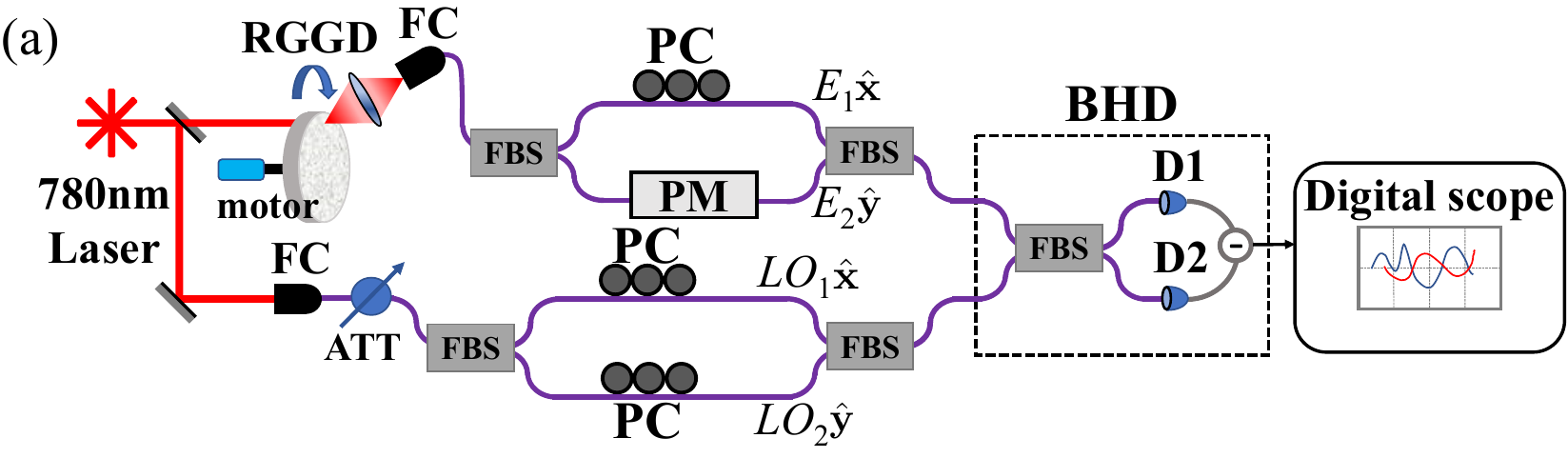}
\includegraphics[width=7cm]{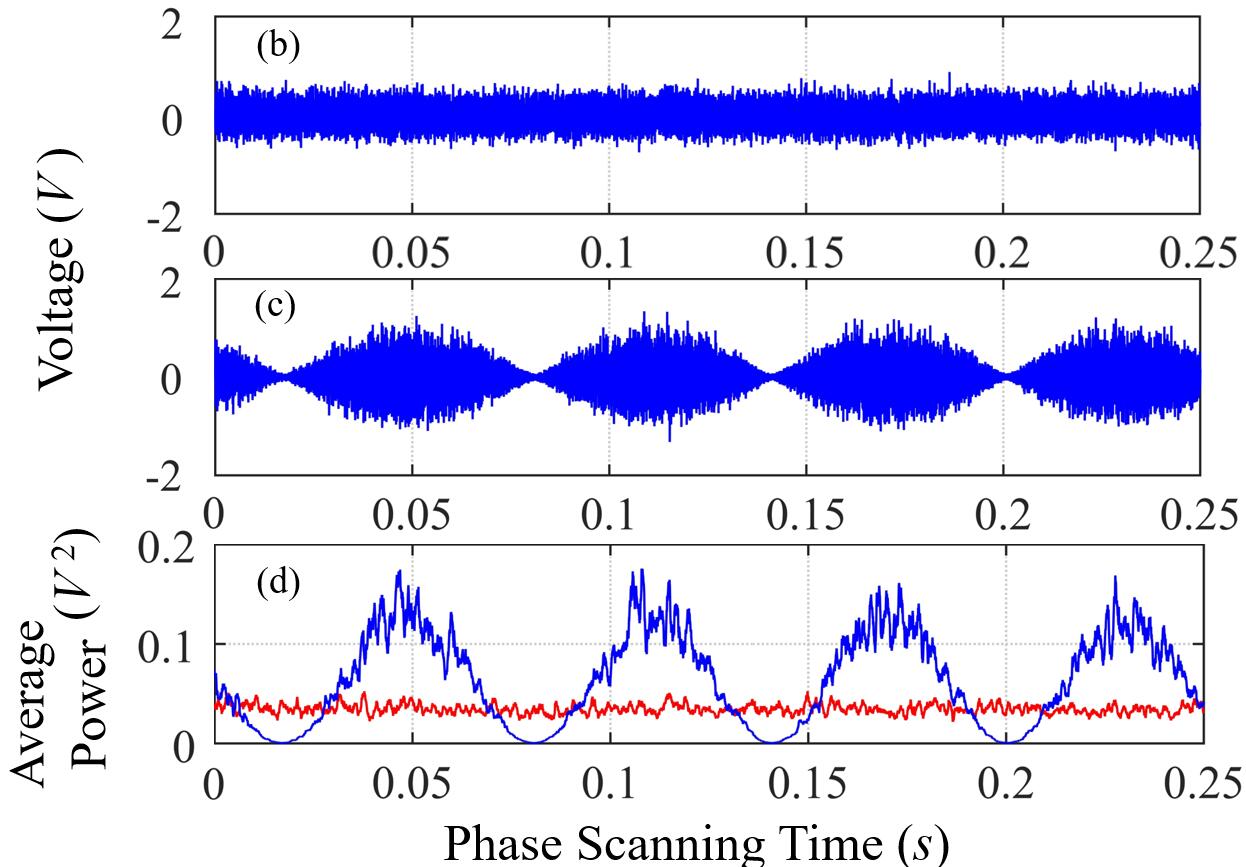}
	\caption{(a) Experimental setup for interference between CW polarization orthogonal fields by homodyne detection. RGGD: rotating ground glass disc.
(b) Raw experimental data of the photo-current from balanced homodyne detector when one arm is blocked and (c) when two arms are open for interference.
(d) Current power $\langle i_{BHD}^2\rangle$  of the output from balance homodyne detection as a function of phase scan in time for the polarization interferometer. The red trace is for one arm with no interference; blue trace is for two arms open.}
	\label{exp-cl}
\end{figure}

The experiment is performed with cw thermal light fields in orthogonal modes.   The experimental setup is shown in Fig.\ref{exp-cl}(a), where a phase-randomized classical cw thermal light field enters a Mach-Zehnder interferometer but with the polarization of one arm rotated 90 degree so that direct detection shows no interference. Of course, interference shows up when we use a polarizer at 45 degree in front of the detector for common mode projection. Now we will recover interference without the polarizer but by direct amplitude addition through balance homodyne detection (BHD) with two local oscillators that each match the polarization modes of the interfering fields. The classical thermal field is generated from the scattering of a coherent laser by a rotating ground glass disc (RGGD) and has a coherence time of 6 $\mu s$, determined by the speed of rotation and the size and location of the laser spot on the disc. The local oscillators are directly split from the laser output. 

The results of the recovered interference is shown in Fig.\ref{exp-cl}(b-d). We record the output of the balanced homodyne detection (BHD) with a fast digital oscilloscope (bandwidth = 5 GHz, sampling rate = 10Gb/sec) as the phase of the interferometer is slowly scanned (15 Hz). We first  do it when one arm of the interferometer is blocked and then when both arms are open, as shown in Fig.\ref{exp-cl}(b) and (c), respectively. Notice that the output of BHD can be both positive and negative. An interference pattern already shows up in the output current in Fig.\ref{exp-cl}(c). But to see clearly the interference fringe, we take an average of the square of the outputs: $\langle i^2_{BHD}\rangle$. The average is over the fluctuations of the input field and it is done experimentally by taking time average over $T_{av}\gg T_c$:  $\langle i^2_{BHD}\rangle\equiv (1/T_{av})\int_{T_{av}} dt  i^2_{BHD}(t)$. Here, $T_{av} =1000 \mu s \gg T_c =6 \mu s$ for averaging out the fluctuations of the field. The results are shown in Fig.\ref{exp-cl}(d) as the red and blue traces, corresponding to Fig.\ref{exp-cl}(b) and (c), respectively. Blue curve shows clearly the interference fringes as the phase is scanned whereas no interference pattern is shown in the red curve. The absence of interference in the homodyne detection of one arm only indicates that there is no coherence between the input field and the local oscillator, as is the case in the derivation of Eqs.(\ref{HD2-i}).

\section{Variation I: Path-unbalanced interferometer beyond coherence length}

We now apply the general idea discussed in the previous section to different situations. The first variation is a path-unbalanced interferometer.

The recovery of interference in the previous section requires multiple mode matched LOs for the amplitude measurement of interfering fields in different modes. We can relax this requirement for cw fields in the same polarization and use only one LO field. In this case, interference can be recovered in a path-unbalanced interferometer even for path difference far beyond coherence length.

\begin{figure}[t]
\includegraphics[width=7cm]{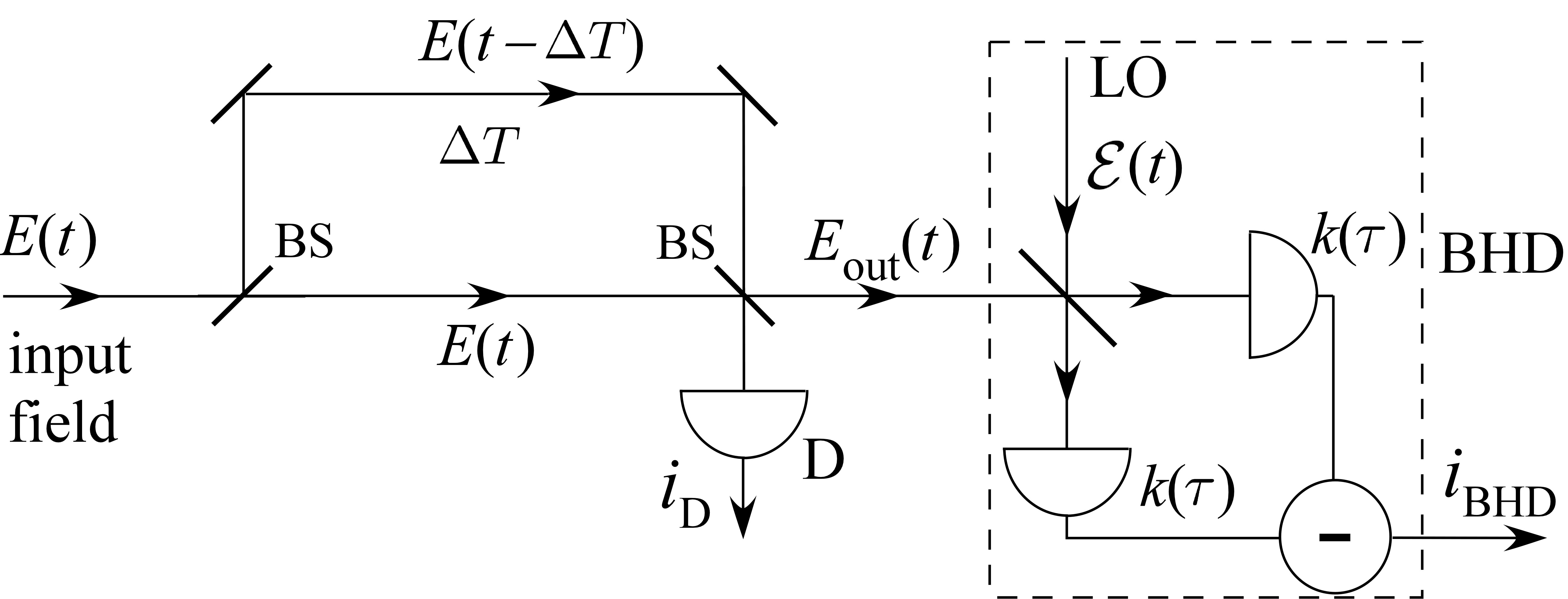}
	\caption{Unbalanced Mach-Zehnder interferometer beyond coherence time.}
	\label{unMZ}
\end{figure}

\vskip 0.1 in 
\noindent {\bf Theory}

Consider a Mach-Zehnder interferometer with a large imbalance $\Delta T$, as shown in Fig.\ref{unMZ}. With 50:50 beam splitters, the output field of the interferometer is related to the input field in the form of
\begin{eqnarray}\label{cw}
E_{out}(t) = [E(t) +  E(t-\Delta T)]/2.
\end{eqnarray}
It is well-known that, if $\Delta T\gg T_c$, the coherence time of the input field, there will be no interference showing up in direct detection regardless of detector's response. Now let us make homodyne detection with an LO of ${\cal E} = |{\cal E}|e^{-i\omega_0 t+i\varphi}$, where $\omega_0$ is the center frequency of input field $E(t)$. With a response function of $k(\tau)$ for the homodyne detector, the output photo-current is
\begin{eqnarray}\label{cw-i}
i_{HD}(t) &\propto & |{\cal E}|\int d\tau k(t-\tau)[X_{\varphi}(\tau) + X_{\varphi}'(\tau-\Delta T)]\cr
&=&|{\cal E}|\int d\tau [k(t-\tau)X_{\varphi}(\tau) \cr
&&\hskip 0.5 in + k(t-\tau-\Delta T) X_{\varphi}'(\tau)],
\end{eqnarray}
where $X_{\varphi}(\tau)= E(\tau)e^{i\omega_0 \tau-i\varphi}+E^*(\tau)e^{-i\omega_0 \tau+i\varphi}$ and $X_{\varphi}'(\tau-\Delta T)= E(\tau-\Delta T)e^{i\omega_0 \tau-i\varphi}+E^*(\tau-\Delta T)e^{-i\omega_0 \tau+i\varphi}$. Notice the amplitude addition in $i_{HD}$ above. With the correlation functions calculated as follows:
\begin{eqnarray}\label{XX}
\langle X_{\varphi}(\tau) X_{\varphi}(\tau')\rangle &=&
\langle X_{\varphi}'(\tau) X_{\varphi}'(\tau')\rangle \cr
&=&  \langle E(\tau) E^*(\tau')\rangle e^{i\omega_0(\tau-\tau') }+ c.c.\cr
&=& I_0\gamma (\tau-\tau')e^{i\omega_0(\tau-\tau') } +c.c.\cr
\langle X_{\varphi}(\tau) X_{\varphi}'(\tau')\rangle &=& I_0 \gamma(\tau-\tau') e^{i\omega_0(\tau-\tau'-\Delta T)} + c.c.\cr
\langle X_{\varphi}'(\tau) X_{\varphi}(\tau')\rangle &=& I_0\gamma(\tau-\tau') e^{i\omega_0(\tau-\tau'+\Delta T)} + c.c. ,~~~~~~
\end{eqnarray}
which show the amplitude correlations ($\gamma$-function) between interfering fields at different times $\tau,\tau'$, we can evaluate the current power $\langle i_{HD}^2(t)\rangle$ with the following result:
\begin{eqnarray}\label{cw-isq}
\langle i_{HD}^2(t)\rangle 
&=& 2I_0 |{\cal E}|^2\int d\tau  d\tau' k(t-\tau) k(t-\tau+\tau') e^{-i\omega_0\tau' } \cr && \times  \big [2\gamma (\tau')+\gamma (\tau'-\Delta T)+\gamma (\tau'+\Delta T)\big] +c.c.\cr &&
\end{eqnarray}

Depending on the response function $k(\tau)$ of the detectors, we have the following scenarios: 

\noindent (i) Detectors with response time $T_R$ (the width of $k(\tau)$) are much slower than the coherence time ($T_c$) of the field: $T_R\gg T_c$. In this case, $k(t-\tau+\tau')$ is nearly unchanged for the range of $\tau'$-integral, which is of the order of $T_c$ determined by the width of $\gamma$-function, and can be pulled out of the $\tau'$-integral. With $\omega_0$ being the central frequency of $E(t)$ field, we can write $\gamma(\tau')=|\gamma(\tau')|e^{-i\omega_0\tau'+i\phi_{\gamma}}$ so that both $|\gamma(\tau')|$ and $\phi_{\gamma}$ are slow functions of $\tau'$. In this way, we define
$\bar T_c \equiv \int d\tau' \gamma(\tau')e^{i\omega_0\tau'}$, which can be shown as a positive quantity for stationary fields and is of the order of the coherence time $T_c$. Hence, we have
\begin{eqnarray}\label{cw-isq2}
&& \langle i_{HD}^2(t)\rangle \propto 4 I_0|{\cal E}|^2 Q_2\bar T_c [1+{\cal V}(\Delta T)\cos(\omega_0 \Delta T)] ~~~~~~
\end{eqnarray}
where $Q_2 \equiv \int d\tau k^2(\tau)$ and
\begin{eqnarray}\label{V}
{\cal V}(\Delta T)\equiv \frac{\int d\tau k(\tau)k(\tau+\Delta T)}{\int d\tau k^2(\tau)}.
\end{eqnarray}
This gives interference as long as $T_R \sim \Delta T$ with a visibility ${\cal V}(\Delta T)\rightarrow 1$ if $T_R\gg \Delta T$, which can always be satisfied if we take long time average of the current before taking the power measurement. In this case, we guarantee the superposition of the amplitudes of the fields from the two paths even though they are very off balance with $\Delta T\gg T_c$. The result is independent of $T_c$. 

{ But if $T_R\ll \Delta T$, ${\cal V}=0$ and no interference shows up. This is because in this case, the detectors are fast enough to resolve the amplitudes of the two paths of the interferometer and the resolved amplitudes have no phase relation because they are outside of the coherence time range. However, the photo-current still contains the information of the amplitudes of the input field in the split fields in the two paths, which are just separated in time by $\Delta T$. So, we can recover interference by constructing $i_{+}= i_{HD}(t)+i_{HD}(t+\Delta T_e)$ with an electronic delay $\Delta T_e$. For $\Delta T_e\sim \Delta T \gg T_c$, by following the same procedure as before, it is straightforward to show
\begin{eqnarray}\label{cw-ipm-sq0}
&&\langle i^2_{+}(t) \rangle \propto 8I_0|{\cal E}|^2Q_2\bar T_c\cr 
&&\hskip 0.6 in \times \Big[1 + \frac{1}{2}{\cal V}(\Delta T-\Delta T_e)|\cos(\omega_0 \Delta T)\Big].~~~~~~
\end{eqnarray}
The large imbalance $\Delta T$ can be compensated by the electronic delay $\Delta T_e$ and interference shows up in $\langle i^2_{+} \rangle$.

\noindent (ii) Detectors ($T_R$) are much faster than the coherence time ($T_c$) of the field: $T_R\ll T_c$. In this case, $k(\tau)$-function acts as a $\delta$-function in the integral and with $Q_1\equiv \int d\tau k(\tau)$, Eq.(\ref{cw-isq}) becomes
\begin{eqnarray}\label{cw-isq3}
&& \langle i_{HD}^2(t)\rangle \approx  4 I_0|{\cal E}|^2 Q_1^2  \cr
&& \hskip 0.7 in \times \big [2\gamma (0)+\gamma (-\Delta T)+\gamma (\Delta T)\big] +c.c.~~~~
\end{eqnarray}
which shows no interference because $\gamma (-\Delta T)=0=\gamma (\Delta T)$ for the unbalanced case of $\Delta T\gg T_c$. This is similar to the case of direct intensity measurement.  However, we can consider $i_{+}= i_{HD}(t)+i_{HD}(t+\Delta T_e)$ as before. To see it better, we go back to Eq.(\ref{cw-i}) for the photo-current and with $k(\tau)$ as a $\delta$-function, we have
\begin{eqnarray}\label{cw-i-Q}
i_{HD}(t) \propto Q_1|{\cal E}|[X_{\varphi}(t) + X_{\varphi}'(t-\Delta T)].
\end{eqnarray}
This clearly shows the addition of amplitudes at different times. Obviously, no interference shows up in $\langle  i^2_{HD}\rangle$ if $\Delta T\gg T_c$. But for  $i_{+}= i_{HD}(t)+i_{HD}(t+\Delta T_e)$, we have 
\begin{eqnarray}\label{cw-ipm}
&&i_{+}(t) \propto Q_1|{\cal E}|[X_{\varphi}(t) + X_{\varphi}'(t-\Delta T)\cr &&\hskip 0.5in +X_{\varphi}(t+\Delta T_e) + X_{\varphi}'(t-\Delta T+\Delta  T_e)].~~~~~~
\end{eqnarray}
With $\Delta T \sim \Delta T_e\gg T_c$ and $\langle E(t)E(t')\rangle = 0$, it is straightforward to find
\begin{eqnarray}\label{cw-ipm-sq}
\langle i^2_{+}(t) \rangle \propto 8Q_1^2|{\cal E}|^2I_0\Big[1 + \frac{1}{2}|\gamma(\Delta T-\Delta T_e)|\cos(\omega_0 \Delta T)\Big].~~~~~~
\end{eqnarray}
This shows interference with maximum visibility of 1/2 when $\Delta T=\Delta T_e$. 50\% instead of 100\% maximum visibility is because of the non-overlap of the middle two terms in Eq.(\ref{cw-ipm}). }

\noindent (iii) For the intermediate case of $T_R\sim T_c \ll \Delta T$, it is straightforward to see from Eq.(\ref{cw-isq}) that the contribution from the second and third terms is zero so that $\langle i_{HD}^2\rangle$ is independent of the path difference $\Delta T$, showing no interference.  Then, similarly as before, we can consider $i_+ = i_{HD}(t) + i_{HD}(t+\Delta T_e)$ with $\Delta T_e\sim \Delta T$. By following the procedure leading to Eq.(\ref{cw-ipm-sq0}), we can show
\begin{eqnarray}\label{cw-ipm-sq3}
&&\langle i^2_{+}(t) \rangle \propto 8I_0|{\cal E}|^2K_2\cr 
&&\hskip 0.6 in \times \Big[1 + \frac{1}{2}{\cal V}'(\Delta T-\Delta T_e)\cos(\omega_0 \Delta T+\theta_0)\Big],~~~~~~
\end{eqnarray}
where 
\begin{eqnarray}\label{K2}
K_2 &\equiv &\int d\tau d\bar \tau k(\tau)k(\tau+\bar \tau)\gamma(\bar \tau)e^{i\omega_0\bar\tau} \cr
&=& \int d\omega |k(\omega)|^2 S(\omega_0-\omega)  > 0
\end{eqnarray}
and 
\begin{eqnarray}\label{V-prime}
{\cal V}' e^{i\theta_0} &\equiv &\int d\tau d\bar \tau k(\tau)k(\tau+\bar \tau+ \Delta T_e-\Delta T)\gamma(\bar \tau)e^{i\omega_0\bar\tau} \cr
&=& \int d\omega |k(\omega)|^2 S(\omega_0-\omega)  e^{i\omega(\Delta T_e - \Delta T)} ,
\end{eqnarray}
where $k(\omega), S(\omega)$ are the Fourier transformation of $k(\tau), \gamma(\tau)$, respectively and $S(\omega)$ is the spectrum of the field and thus is always positive.

\vskip 0.2 in
\noindent {\bf Experimental verification}

\begin{figure}[t]
\includegraphics[width=8.5cm]{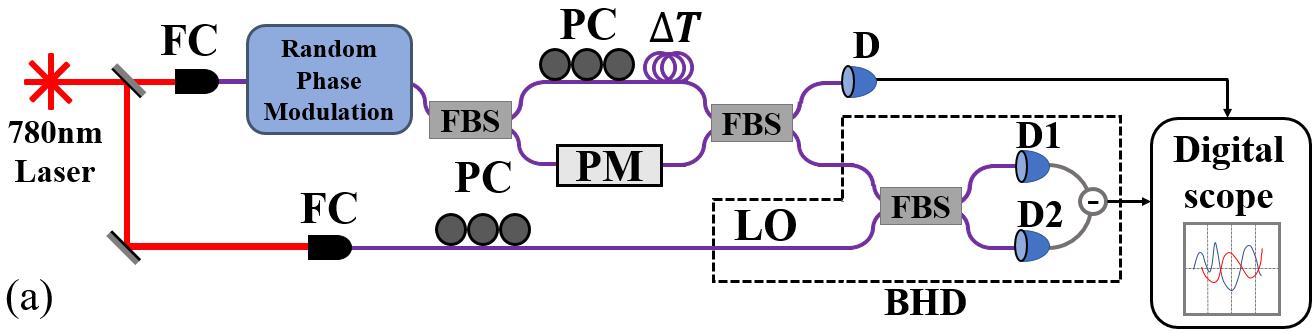}
\includegraphics[width=8cm]{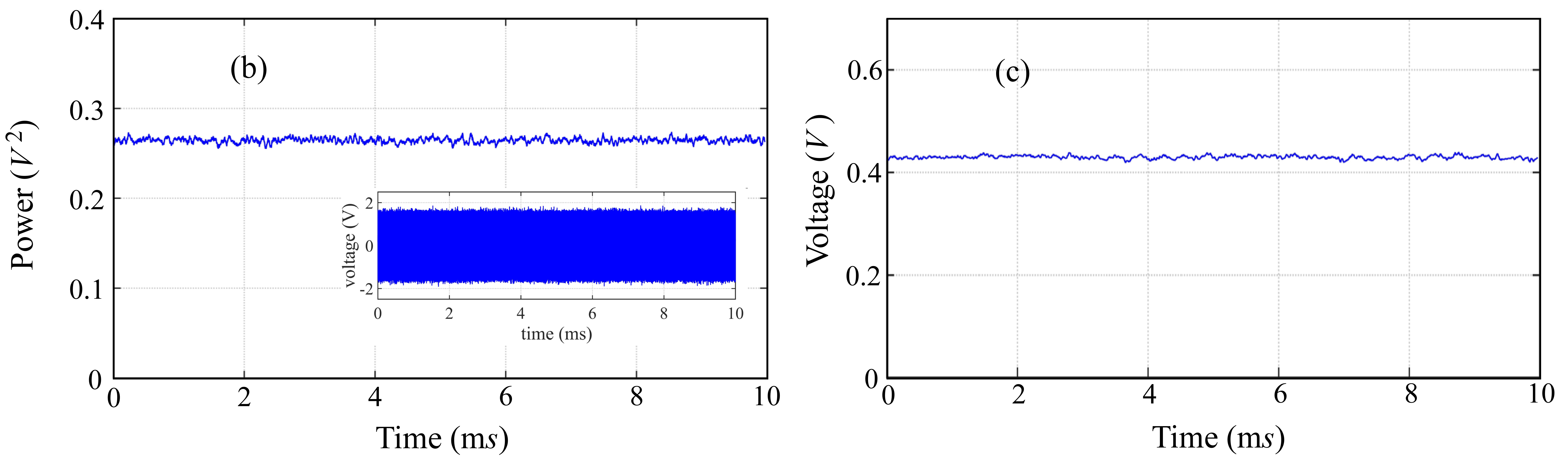}
	\caption{(a) Schematics for unbalanced Mach-Zehnder interferometer beyond coherence time with a phase-randomized field of coherence time $T_c= 2 ns$. PC: polarization controller; FC: fiber coupler; FBS: fiber beam splitter; LO: local oscillator field; BHD: balanced homodyne detector. (b) The average power $\langle i_e^2\rangle$ of fast RF signal from the balanced homodyne detectors ($T_{av} = 31.25\mu s$); Inset: raw data of $i_e(t)$; (c) Direct intensity measurement by D3. All shows no interference.}
	\label{unb-cw}
\end{figure}

We next demonstrate experimentally how to recover interference in an unbalanced interferometer with path difference beyond coherence length of the input field. The schematics is shown in Fig.\ref{unb-cw}(a). The input field to the interferometer is a phase-randomized field generated by passing a coherent state from laser through an electric-optic phase modulator driven by a white noise source of 500MHz bandwidth, resulting in a coherence time of 2 ns.  The input field is injected to a fiber ($n=1.5$) Mach-Zehnder interferometer with a path imbalance of $\Delta L =14 m (\Delta T = n\Delta L/c = 63 ns > T_c =  2 ns)$. Polarization of one arm is adjusted by a polarization controller (PC) for optimum polarization mode match while the phase of the other arm is modulated (PM). Balanced homodyne detection (BHD) is performed at one of the outputs while the intensity measurement by a slow detector D3 ($<$ 1MHz) is done on the other. The local oscillator for BHD is split from the laser. The fast RF photo-current from BHD (up to 150 MHz) is recorded by a digital scope. Direct intensity measurement current by D3 is also recorded. 

\begin{figure}[t]
\includegraphics[width=6.5cm]{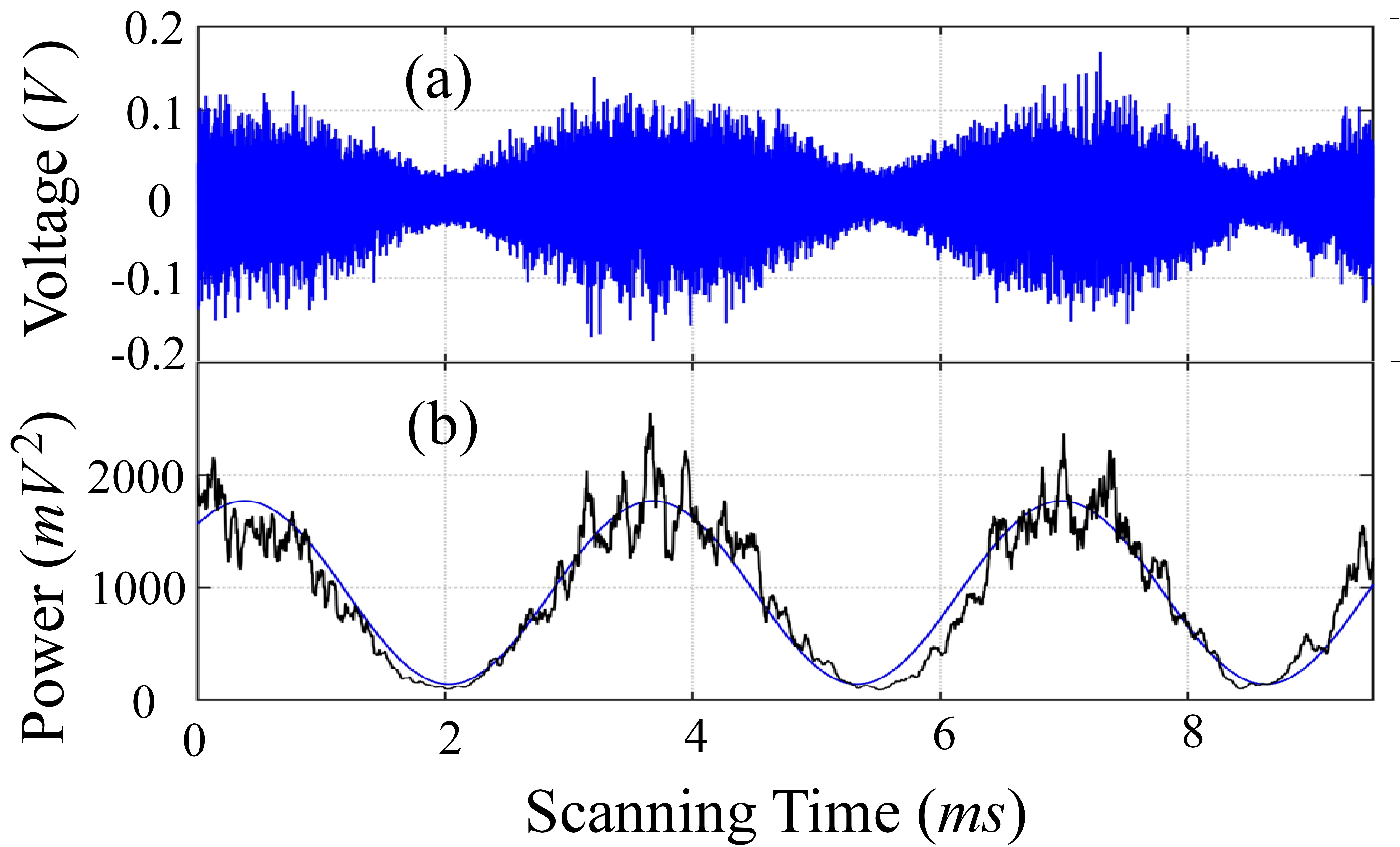}
\includegraphics[width=8.5cm]{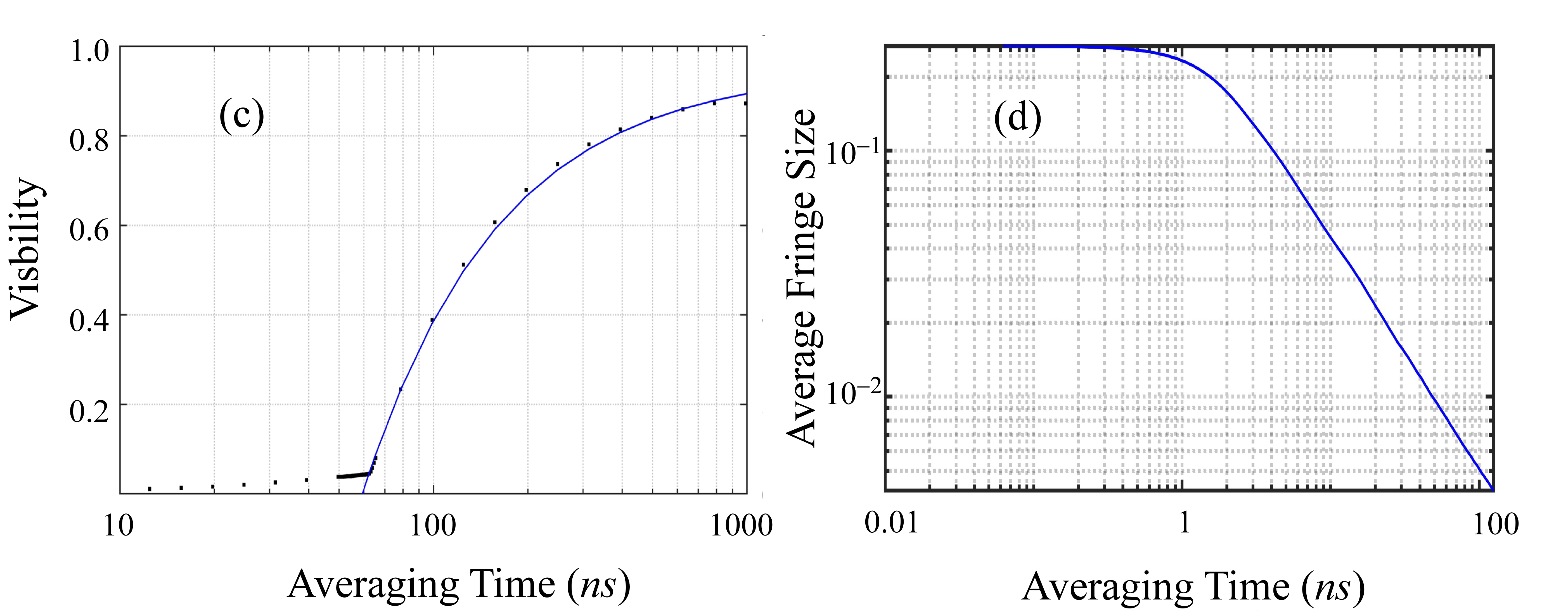}
	\caption{Recovery of interference by time average: (a) Time-averaged photo-current $\bar i_e(t)$ with time-averaging duration $T=625 ns >$ imbalance $\Delta T = 63 ns$ of the unbalanced interferometer. (b) $\langle \bar i_e^2 \rangle$ ($T_{av} = 62.5\mu s$), showing interference fringe with a visibility of 85 \%.  (c) Visibility as a function of the time-averaging duration $T$. The solid curve is a fit to ${\cal V}(T) = 1-\Delta T/T$ for $T>\Delta T$. { (d) The average size of fringe (mid level of $\langle \bar i_e^2 \rangle$) as a function of time-averaging duration $T$.}}
	\label{unb-cw2}
\end{figure}

The data of the output current power average $\langle i_e^2\rangle$ ($T_{av} = 62.5\mu s$) of the fast RF signal from the balanced homodyne detectors (BHD) as well as the direct intensity measurement (D3) are presented in Fig.\ref{unb-cw}(b,c), showing no interference fringe because the interferometer path difference is bigger than the coherence length of the input field. Raw data of $i_e(t)$ is shown in the inset of Fig.\ref{unb-cw}(b). On the other hand, if we take a time-average of the fast current: $\bar i_e(t) = \frac{1}{T}\int_{t}^{t+T} d \tau i_e(\tau)$, for large $T (\gg \Delta T)$, we find an interference pattern in $\bar i_e(t)$ and $\langle \bar i_e^2(t)\rangle$ ($T_{av} = 62.5\mu s$), as shown in Fig.\ref{unb-cw2}(a,b). It turns out the visibility of the interference pattern depends on the time average period $T$, as shown in Fig.\ref{unb-cw2}(c). In fact, in this case, the time average is equivalent to a response function of $k(\tau) = 1/T$ for $0 < \tau < T$ but is zero otherwise. The solid curve in Fig.\ref{unb-cw2}(c) is from Eq.(\ref{V}), where we find ${\cal V}(T) = 1-\Delta T/T$ for $T>\Delta T$ but is zero for $T<\Delta T$, giving a cut-off time of $\Delta T$. 

 Although the long averaging period increases the visibility, it will reduce the size of $\bar i_e$ and $\langle\bar i_e^2\rangle$, i.e., the size of the interference fringe if $T>T_c$. The reason is that $\bar i_e$ is a time average of $X(\phi)$-quadrature so it averages out the phase fluctuations if $T>T_c$. When $T<T_c$, the detector or $\bar i_e$ will see the phase fluctuations of the input field and $\bar i_e$ will not drop too much. But when $T>T_c$, $\bar i_e$ will average out the phase fluctuations and thus it will decrease for $T> T_c$ and drops to zero for $T\gg T_c$. We plot the average of $\langle\bar i_e^2\rangle$ (mid-level of the fringe) as a function of $T$, as shown in Fig.\ref{unb-cw2}(d). It starts to drop at $T\sim T_c (\sim 1 ns)$ and goes as $1/T$, as predicted from the qualitative argument above. We can also see this from another point of view as follows:
The time average is equivalent to a narrow band filter on the photo-current. So, it is not surprise to recover interference. But this reduces the bandwidth of the signal and cuts off the high frequency components thus reduces the signal size  as we see in Fig.\ref{unb-cw2}(d). When $T<T_c$ or bandwidth of filter $=1/T > 1/T_c =$ bandwidth of field, no reduction of the signal, as discussed earlier. But when the filter bandwidth ($1/T$) is smaller than the field bandwidth ($1/T_c$), or $T>T_c$, the filter will cut the signal level. 

\begin{figure}[t]
\includegraphics[width=8cm]{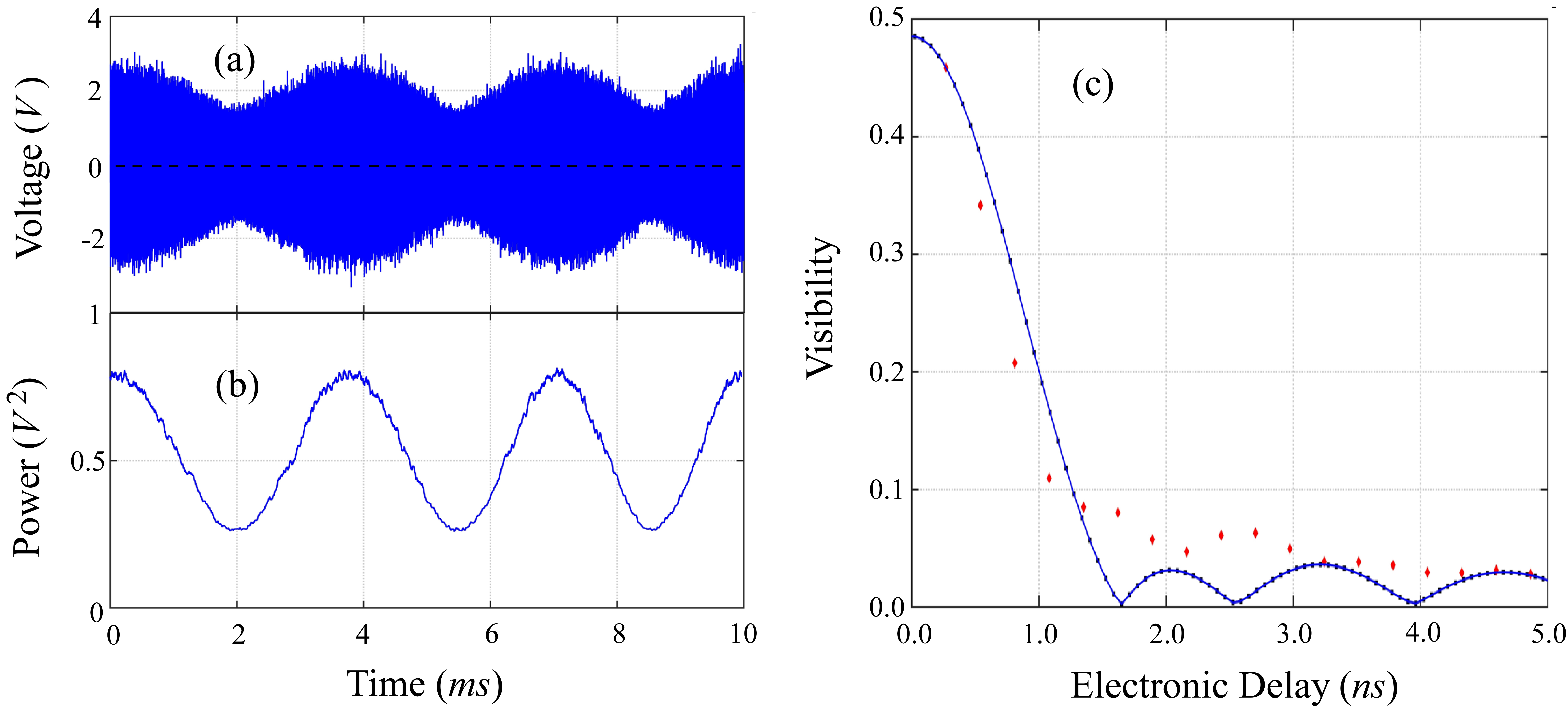}
	\caption{(a) Raw data of $i_+(t) = i_e(t) + i_e(t+\Delta T_e)$ with $\Delta T_e = 63 ns =\Delta T$. (b) The power $\langle i_+^2 \rangle$ of $i_+(t)$ ($T_{av} = 62.5\mu s$), showing interference. (c) Visibility as a function of the electronic compensation time delay $\Delta T_e -\Delta T$ (blue solid curve). The red dots are values numerically calculated from Eq.(\ref{V-prime}).}
	\label{unb-i+}
\end{figure}

To circumvent the above, we can also recover the interference pattern through direct amplitude addition by constructing $i_+ = i_e(t) + i_e(t+\Delta T_e)$ with an electronic delay $\Delta T_e$, as predicted by Eq.(\ref{cw-ipm-sq}). When $\Delta T_e \sim \Delta T$, we find $i_+$ shows interference pattern as demonstrated in the raw data in Fig.\ref{unb-i+}(a) and in the average power $\langle i_+^2\rangle$ in Fig.\ref{unb-i+}(b). The extracted visibility is shown as a function of electronic compensation $\Delta T_e-\Delta T$ in Fig.\ref{unb-i+}(c) (blue curve). The red dots in Fig.\ref{unb-i+}(c) are obtained from numerical calculation based on Eq.(\ref{V-prime}) after Fourier transformation of the detector's response function and the spectrum of the input, which can both be measured with an electronic spectral analyzer in the experiment. Only a discrete number of simulated values are shown because of the long integration time in the numerical calculation. The agreement between the experimental data (blue curve) and theoretical prediction (red dots) is quite good.  But notice that the experimental data (blue curve) show a number of zeros for the visibility at certain delays. We do not know the reason at present.

Although no coherence between the input signal field and the LO field is required to reveal the interference, the main role of the LO field is to set a stable phase reference for the two interfering fields to compare. So, its coherence time $T_{LO}$ must be longer than all time scales in the experiment: $T_{LO} \gg T_c, T_R, \Delta T$. 

\section{Variation II: Pulsed and Quasi-cw cases}

\subsection{Interference of pulsed fields in well distinguishable states of both polarization orthogonal and temporal non-overlapping modes}

\noindent {\bf Theory}

\begin{figure}[t]
\includegraphics[width=8.5cm]{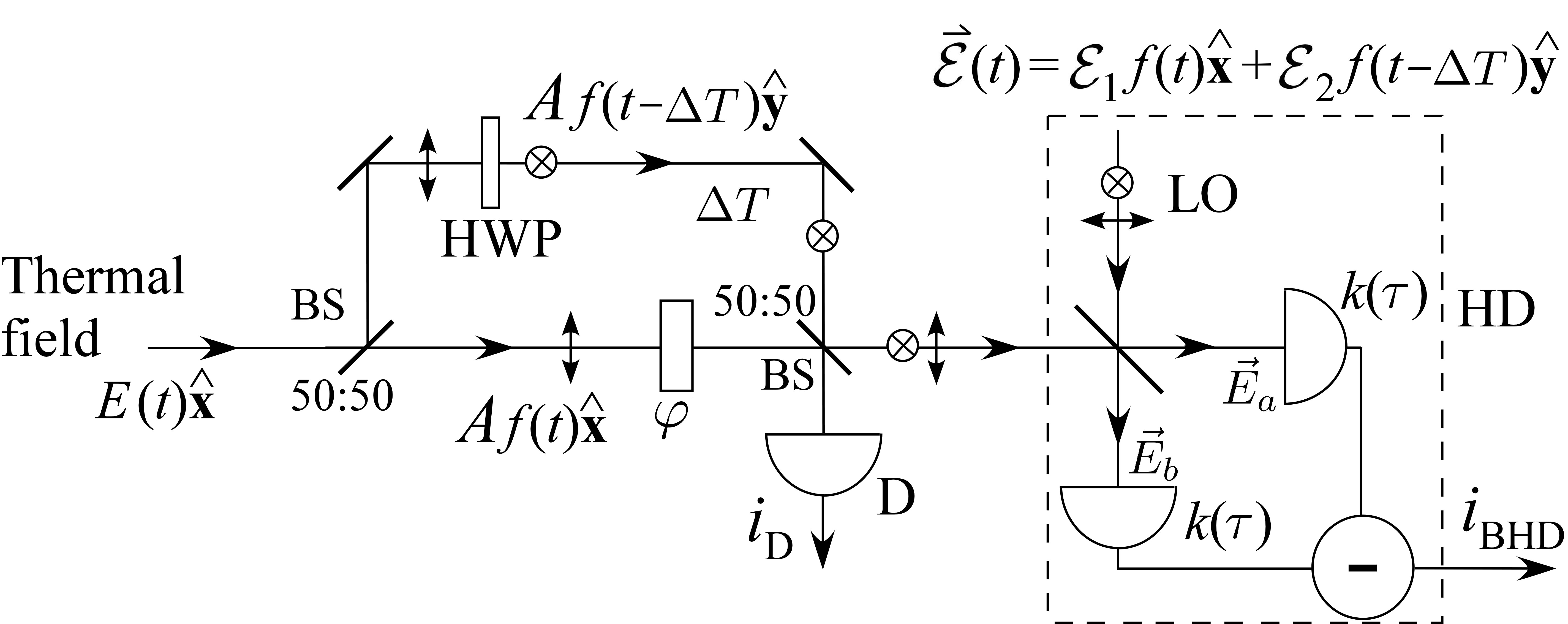}
	\caption{Interference between two orthogonally polarized and temporally non-overlapping fields.}
	\label{XY2}
\end{figure}

Although the discussion in Sec.II uses two LOs with orthogonal polarizations of $\hat {\bf x}$ and $\hat {\bf y}$, it can be viewed as one LO but with a polarization of $\hat \epsilon = \hat {\bf x}\cos \theta + e^{i\delta}\hat {\bf y}\sin \theta$ with $\tan\theta \equiv |
{\cal E}_1/{\cal E}_2|$ and $\delta \equiv \varphi_2-\varphi_1$. For the case of unbalanced interferometer in Sec.III, one may associate the phenomenon with narrow band filtering often used in homodyne detection \cite{wolf,agar,mandel93,rauch} even though it is operated here in time domain without any filtering.  To demonstrate the general interference recovery technique discussed in Sec II in a more dramatic way, now let us add temporal mode to completely separate the two fields. This corresponds to the non-stationary case of a short pulsed input field $E(t) = A f(t)e^{-i\omega_0 t}$ entering an unbalanced Mach-Zehnder interferometer as shown in Fig.\ref{XY2}. No interference occurs in direct detection  even if we use a polarizer oriented at 45 degree for projection because of the temporal distinguishability due to delay $\Delta T$ much larger than the width of the pulses profile $f(t)$. Here, $A$ is a slowly varying envelope of the field, carrying information about field fluctuations (coherence). Now we consider homodyne detection. But in order to have contribution from both arms, we need to have mode matched LOs in homodyne detection:
\begin{eqnarray}\label{ELO-p}
\vec {\cal E}(t) = [{\cal E}_1f(t)\hat {\bf x}+ {\cal E}_2f(t-\Delta T)\hat {\bf y}]e^{-i\omega_0 t}
\end{eqnarray}
with ${\cal E}_j=|{\cal E}_j|e^{i\varphi_j} (j=1,2)$. Here, in order to demonstrate the dramatic effect, we also assume the two fields have orthogonal polarizations, which do not matter in amplitude addition. The detectors are normally slow compared with the pulse profile function $f(t)$. So, the instantaneous output current from HD is a time integral over the pulse width:
\begin{eqnarray}\label{HD-pls}
i_{\rm pulse}(t) &\propto & \int d\tau k(t+\tau)\Big\{|f(\tau)|^2|{\cal E}_1| X_1(\varphi_1)\cr
 &&\hskip 0.5in + |f(\tau-\Delta T)|^2|{\cal E}_2|X_2(\varphi_2)\Big\}\cr
&\approx &   k(t)|{\cal E}_1| X_1(\varphi_1)+ k(t+\Delta T)|{\cal E}_2| X_2(\varphi_2),~~~~
\end{eqnarray}
where $k(\tau)$ is the response function of the detectors, and $X_1(\varphi_1)\equiv A(t) e^{-j\varphi_1} + A^*(t)e^{j\varphi_1}$, $X_2(\varphi_2)\equiv A(t-\Delta T) e^{-j(\varphi_2+\omega_0\Delta T)} + A^*(t-\Delta T)e^{j(\varphi_2+\omega_0\Delta T)}$. In deriving the above equation, we used $\int dt f(t)f(t-\Delta T)=0$ for $\Delta T \gg$ the width of $f(t)$. The overall output of homodyne detection for the pulsed case is the current power in a time integral over the response function and an average over the fluctuations of $A(t)$:
\begin{eqnarray}\label{HD-pls2}
&&\int i^2_{\rm pulse}(t) dt \propto 2\langle |A(t)|^2\rangle Q_2 (|{\cal E}_1|^2+|{\cal E}_2|^2)\cr && \hskip 1 in \times [1+{\cal V}_{\rm p} \cos(\phi_{\gamma}+\Delta \varphi -\omega_0\Delta T)],~~~~
\end{eqnarray}
with $Q_2\equiv \int dt k^2(t)$, $\Delta \varphi \equiv \varphi_1-\varphi_2$, and
\begin{eqnarray}\label{V-p}
{\cal V}_{\rm p} \equiv  \frac{2|\gamma(\Delta T)|\sqrt{\lambda}}{1+\lambda}\frac{\int dt k(t)k(t+\Delta T)}{\int dt k^2(t)},
\end{eqnarray}
where $\lambda \equiv |{\cal E}_1|^2/ |{\cal E}_2|^2, \gamma(\tau)\equiv \langle A(t)A^*(t-\tau)\rangle/\langle |A(t)|^2\rangle$ and $\phi_{\gamma}\equiv {\rm Arg}(\gamma)$. In deriving Eq.(\ref{HD-pls2}), we assumed $A(t)$ has a random phase fluctuation so that $\langle A(t)\rangle = 0$ and $\langle A^2(t)\rangle = 0$.

Similar to Eq.(\ref{HD2-i}), the interference patterns show up in the output of homodyne detection. The difference is in the dependence of visibility on the response function $k(\tau)$ in the pulsed case. From Eq.(\ref{HD-pls}) we can see the role played by the response function of the detectors. Both Eqs.(\ref{HD2}) and (\ref{HD-pls}) show the addition of the amplitude, although we need to assume the slowness of the detectors or $k(t)\approx k(t+\Delta T)$ so the amplitudes at different time can be added to produce interference. On the other hand, if detectors are fast so that they can resolve the two pulses separated by $\Delta T$, that is $k(t)k(t+\Delta T)=0$, the two amplitudes will not be added at time $t$ and ${\cal V}_{\rm p} =0$, resulting in no interference.  But we can make post-detection data processing to recover interference, as will be seen next.

\begin{figure}[t]
\includegraphics[width=8cm]{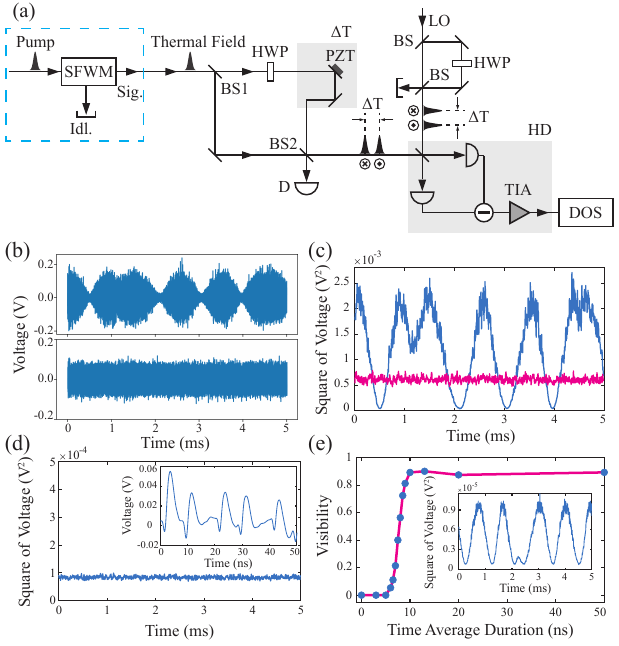}
	\caption{(a) The experimental setup of for interference between two pulses distinguished in both temporal and polarization modes. BS, 50/50 beam splitter; SFWM, spontaneous four wave mixing; HWP, half wave plate, PD, power detector; HD, homodyne detection. (b, c) are obtained by HD with response time $T_R$ of about 100 ns. The upper and lower traces in (b) correspond the photo-current $\langle i_{HD}(t)\rangle$ when both arms and only one arm of the interferometer are unblocked, respectively; the blue and pink traces in (c) are envelop of $\langle i_{HD}^{2}(t) \rangle$ for the upper and lower traces in (b). (d,e) are obtained by HD with response time $T_R$ of about 4 ns. Plot (d) is the envelop of $\langle i_{HD}^{2}(t) \rangle$ for the unbalanced interferometer with both arms unblocked. Plot (e) is the visibility of interference as a function of the time-averaging duration. The insets in (d) and (e) are the raw data of fast HD and recovery of interference obtained by time average $T=10$ ns, respectively}
	\label{exp-pulse2LO}
\end{figure}

\vskip 0.1 in
\noindent {\bf Experimental verification}  

We can demonstrate experimentally the case of orthogonal temporal fields in Fig.\ref{XY2} with a pulsed thermal field generated by Raman scattering of a pulsed coherent source. The schematics with a train of pulsed thermal field as the input of unbalanced interferometer is shown in Fig.\ref{exp-pulse2LO}(a). The thermal field $E(t)$ having pulse duration and coherence time of about 9 and 8 ps, respectively, is originated from the individual signal field of spontaneous four wave mixing (SFWM) in a pulse-pumped single mode fiber \cite{Ma-pra11}. The thermal state has well defined polarization, and the separation between two adjacent pulses of thermal field, determined by the repetition rate of pump laser, is about 20 ns. The unbalanced interferometer consists of two 50/50 beam splitters (BS), and the path length difference-induced delay between two arms $\Delta T$ is about 7.5 ns. Moreover, we insert a half-wave plate (HWP1) in the delayed arm and rotate its polarization by 90 degree. So the two pulsed fields combined by BS2 are well distinguished in both temporal and polarization modes.	

The two output ports of the unbalanced interferometer are respectively measured by a power detector (PD) and a balanced homodyne detector (HD). To match the modes of fields being measured, we create double LO by sending laser pulses synchronized with the pump of SFWM into an unbalanced Mach-Zehnder (MZ) interferometer, in which the delay is the same as $\Delta T$ and an HWP is inserted in the delayed arm. Under this condition, the homodyne detection measures the amplitudes of two distinguished pulses and adds them up as long as the response time of HD is longer than the delay $\Delta T$. To conveniently extract the results of balanced HD measurement, the differential photo-current of two photo diodes is converted to voltage by a transimpedance amplifier (TIA). The output of HD is recorded by a digital oscilloscope (DOS), in which the sampling rate of 10 GHz for 2 ms contains the results of $10^5$ optical pulses. When the relative phase between two arms is scanned by driving the PZT with a ramping voltage at a rate of about 200 Hz, no interference is observed by the PD. However, interference can be observed from the HD measurement $ i_{HD}(t)$, as shown by the envelop of the upper trace in Fig. \ref{exp-pulse2LO}(b). The lower trace in Fig. \ref{exp-pulse2LO}(b) corresponds to results of HD by blocking one arm of unbalanced interferometer. For clarity, we take average of the square of the data in Fig. \ref{exp-pulse2LO}(b) and obtain $\langle i_{HD}^{2} (t)\rangle$. As shown in Fig. \ref{exp-pulse2LO}(c), the visibility of interference (blue trace) is about 95 $\%$, and the height of the pink trace, which corresponds to one arm measurement, is about $\frac{1}{4}$ of the peaks of interference fringe of the blue trace. 

To illustrate the influence of the response time of HD ($T_R $) upon the visibility of interference as given in Eq. (\ref{V-p}), we replace the HD detectors that have a slow response time of $T_R  \approx 100$ ns with a fast one of $T_R  \approx 4$ ns and repeat the measurement. As shown by the inset in Fig. \ref{exp-pulse2LO}(d), the fast photo-current pulses are able to distinguish the delay $\Delta T$ of about 7.5 ns. In the data processing process, we extract the peak of each pulse and correct it by subtracting the mean value for all the peaks of the $10^5$ pulses. We further analyze the photo-current represented by these peaks by taking square of peak values and plotting the envelop in Fig. \ref{exp-pulse2LO}(d), from which we cannot observe interference, as predicted by Eq.(\ref{V-p}). But when we take a time-average of the fast photo-current pulses, we are able to recover the interference for $T>\Delta T$ (see the inset in Fig. \ref{exp-pulse2LO}(e)). This is because the time average with duration $T$ is equivalent to a square response function of $k(\tau)\approx 1/T$ for $\tau<T$ but $=0$ for $\tau \ge T$. For clarity, we plot the visibility of recovered interference as a function of the time-averaging duration $T$, as shown in Fig. \ref{exp-pulse2LO}(e). It is clear, the visibility increases with time average duration $T$. The result qualitatively agree with Eq. (\ref{V-p}), but the variation trend does not agree (see Fig.\ref{unb-cw2}(c)) because the response function of the fast HD does not satisfy the assumption leading to the solid curve there (Fig.\ref{unb-cw2}(c)). 

\subsection{Interference of quasi-cw pulsed fields with path-imbalance beyond coherence length}

The scheme in previous section requires two LO fields because the interfering fields have two orthoganal polarizations. Consider now a quasi-CW field of a pulse train as the input to the unbalanced interferometer. This situation often occurs for remote sensing/LIDAR application where we send out a well-defined laser pulse trian and detect the information encoded in the returned pulses. So, the temporal profile is known for the returned signal. The LO fields and the probing laser pulse can be different lasers due to different requirement on them but must be synchronized.

Since the interfering fields have the same polarization, only one LO field with a matched pulse train is needed for homodyne.   In this case, the results are similar to the cw case discussed in Sec. III. But the treatment is different. Due to complexity, we leave the details of the derivation to Appendix B and only present the results here.

The input  field of quasi-CW pulse train is described as
\begin{eqnarray}\label{qcw}
E(t) = \sum_{j=-\infty}^{+\infty}  A_j f(t-jT_p)e^{-i\omega_0  (t-jT_p)}
\end{eqnarray}
with $f(t)$ as the normalized pulse profile ($\int dt f^2(t)=1$) and $T_p$ as the separation between adjacent pulses. The width of the pulse ($f(t)$) is typically much shorter than the pulse separation $T_p$. $A_j$ is a complex random variable characterizing the amplitude and phase fluctuations of each individual pulse. We assume it is not related to the LO field.

In the first case of short coherence time for the source, we assume $\{A_j\}$ are independent random variables so that $\langle A_jA_k^*\rangle = I_j\delta_{jk}$, which corresponds to the case of $T_c\ll T_R$ in cw case. Let the imbalance $\Delta T$ of the interferometer be exactly the whole multiple of pulse separation $T_p$: $\Delta T= nT_p$ so that the pulses are still fully overlapped at the second beam splitter (BS2). 

With a response function $k(\tau)$ for the detector and a phase difference of $\theta$ between two arms, the photo-current from direct intensity measurement of the output of the interferometer can be calculated as
\begin{eqnarray}\label{qcw-iD2r}
\langle i_{D}(t)\rangle & \propto & \int d\tau k(t-\tau)
\sum_j f^2(\tau-jT_p) I_j(1+\delta_{n,0}\cos\theta),\cr
&&
\end{eqnarray}
which, as expected, shows no interference unless $n = 0$ (balanced case) irrespective of the response of the detector, even though there is overlap of pulses. 

For the measurement at interferometer output by homodyne detection with LO as ${\cal E}(t) = {\cal E}\sum_j f(t-jT_p)e^{-i\omega_0 (t-jT_p)}$ which matches the pulse profile as the input and has a fixed amplitude: ${\cal E}=|{\cal E}|e^{i\varphi}$, with detectors unable to resolve pulse profile of $f(t)$, the photo-current output is
\begin{eqnarray}\label{qcw-i2r}
i_{HDqcw}(t) & \approx &
|{\cal E}|\sum_j k(t-jT_p) [X_j(\varphi)+X_{j-n}(\varphi-\theta)]\cr
&=& |{\cal E}|\sum_j [k(t-jT_p) X_j(\varphi)\cr
&&\hskip 0.3in + k(t-nT_p-jT_p)X_{j}(\varphi-\theta)],
\end{eqnarray}
which gives amplitude addition.
To see interference, we measure the average current power $\langle i^2_{HDqcw}\rangle$ and after integration over many pulses, we have
\begin{eqnarray}\label{qcw-i2pr}
\int dt \langle i^2_{HDqcw}\rangle \propto
Q_2 |{\cal E}|^2 \langle I_j\rangle [1+ {\cal V}(nT_p)\cos\theta].
\end{eqnarray}
where $Q_2 \equiv \int dt k^2(t)$ and ${\cal V}(nT_p)$ has the same form of Eq.(\ref{V}) as the cw case. Therefore, we recover interference by homodyne measurement even in the unbalanced case beyond pulse coherence ($n\neq 0$) if $nT_p\ll T_R$.

Similar to the cw case, we find that interference disappears or ${\cal V}(nT_p)=0$ if $k(t)$ and $k(t-nT_p)$ do not overlap when imbalance $\Delta T= nT_p\gg T_R$. This is the case when the detector can tell the difference between the two paths of interferometer.  On the other hand, we can still recover interference by constructing $i_+ = i_{HDqcw}(t) + i_{HDqcw}(t+\Delta T_e)$ with an electronic delay $\Delta T_e$ to compensate the optical delay: $\Delta T_e\sim nT_p$, as in the cw case. It is straightforward to show that
\begin{eqnarray}\label{i+sq}
\int dt \langle i^2_+\rangle \propto
Q_2 |{\cal E}|^2 \langle I_j\rangle \Big[1+ \frac{1}{2}{\cal V}(nT_p-\Delta T_e)\cos\theta\Big].~~~~
\end{eqnarray}

Next, in the other extreme case of relatively long coherence time with $T_c\gg T_R$, which usually occurs in remote sensing but with applicable delay $\Delta T$ limited by coherence time $T_c$, we can assume $\langle A_jA_{j+q}^*\rangle = I_q \ne 0$ for $q=-M,...,-1,0,1,2,...,M$ with $M=[T_c/T_p]$ but $\langle A_jA_{j+q}^*\rangle = 0$ for $|q|=M+1, M+2, ...$. For large delay with $\Delta T= nT_p \gg T_c$ or $n \gg M$ beyond the applicable range of traditional remote sensing, we find no interference is present in direct detection because of $\langle A_jA_{j+q}^*\rangle = 0$ for $|q|=M+1, M+2, ...$. Notice that $\langle A_j\rangle =0$ because of the randomness of the input field with respect to the LO field.

However, as we will see, interference can recover in homodyne detection with some manipulation.
The output photo-current from HD is still given by Eq.(\ref{qcw-i2r}). But because of the large delay ($\Delta T = nT_p \gg T_c\gg T_R$), there is no overlap between $k(t-jT_p)$ and $k(t-nT_p-jT_p)$ and 
this still leads no interference (see Appendix B for details). On the other hand,  we can consider the current addition $i_+ = i_{HDqcw}(t) + i_{HDqcw}(t+\Delta T_e)$ with an electronic delay $\Delta T_e$ to compensate the optical delay: $\Delta T_e\sim \Delta T=nT_p$, as before. It has the form of
\begin{eqnarray}\label{qcw-i+}
i_+(t) &\propto &
|{\cal E}| \sum_j [k(t-jT_p) X_j(\varphi)\cr
&&\hskip 0.3in + k(t-nT_p-jT_p)X_{j}(\varphi-\theta)\cr
&&\hskip 0.3in + k(t-jT_p+\Delta T_e) X_j(\varphi)\cr
&&\hskip 0.3in + k(t-jT_p-nT_p+\Delta T_e)X_{j}(\varphi-\theta)],~~~~~~
\end{eqnarray}
and the first and last terms have overlap for interference if $\Delta T_e\sim \Delta T=nT_p$. It can be shown (Appendix B) that the interference is recovered in $i_+$:
\begin{eqnarray}\label{qcw-i+s2r}
\int dt \langle i^2_+\rangle \propto
Q_1^2I_0 |{\cal E}|^2 \Big[1+ \frac{1}{2}(I_m/I_0)\cos\theta\Big],~~~~~~
\end{eqnarray}
where $I_m \equiv \langle A_jA_{j+m}^*\rangle$ with $m = [|\Delta T_e - \Delta T|/T_p]$. This is similar to the cw case of Eq.(\ref{cw-ipm-sq}).

\begin{figure}[t]
\includegraphics[width=8cm]{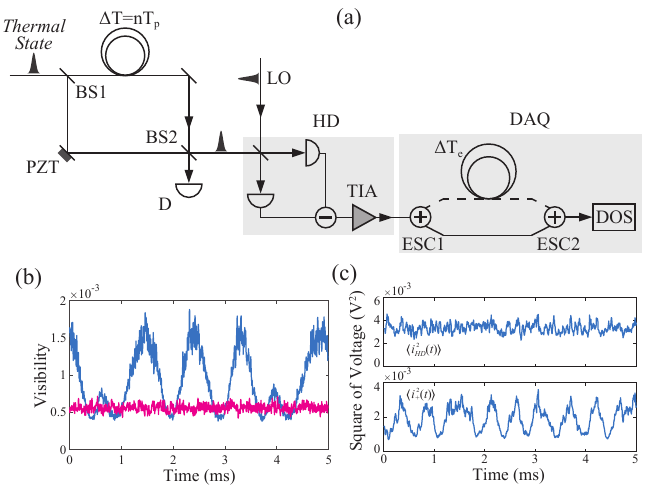}
	\caption{(a)Schematics for unbalanced Mach-Zehnder interferometer with delay $\Delta T= T_P=20$ ns. ESC, electrical splitter circuit; DOS, digital oscilloscope; DAQ, data acquisition system.  (b) The envelop of $\langle i_{HD}^{2}(t) \rangle$  for HD with the detection response time $T_R$ of about 100 ns. The pink trace is when one arm of the interferometer is blocked. (c) The envelop of $\langle i_{HD}^{2}(t) \rangle$ and $\langle i_{+}^{2}(t)\rangle $ for HD with $T_R$ of about 4 ns.}
	\label{exp-pulse2}
\end{figure}

\vskip 0.1 in
\noindent {\bf Experimental verification} 

Now let us perform experiment with pulsed input field to an unbalanced interferometer and recover interference by one HD measurement. Different from two-LO scheme in Fig. \ref{exp-pulse2LO}(a) where the delay of the unbalanced interferometer can be some arbitrary amount, the delay between two arms of unbalanced interferometer must be a multiple of pulse separation, i.e., $\Delta T = nT_{p}$  ($n\geq1 $ is integer). This is because we use one HD with a single pulse train as LO, which needs to overlap with the fields going through both arms of the interferoemeter in order to measure their amplitudes for superposition. For the experimental setup in Fig.\ref{exp-pulse2}(a), the input is the same as that in Fig. \ref{exp-pulse2LO}(a), but the laser pulse train of LO is directly sent into HD to match the mode of field to be measured. In the experiment, we insert a 4.08-m-long ($\Delta L$) standard single-mode fiber (SMF) in one arm of interferometer so that the delay $\Delta T$ is exactly equal to the separation between adjacent pulses, i.e., $\Delta T = n\Delta L/c = 20 ns = T_p$.

When the relative phase between two arms is scanned by PZT, no interference is observed by Detector D placed at one output of the interferometer no matter whether its response time is slow or fast. In the HD measurement of the other output port, we first choose to use detectors of response time $T_R \approx 100$ ns and directly record the output of HD by DOS. The data of DOS is similar to that in Fig.\ref{exp-pulse2LO}(b)-(c). For brevity, we omit the procedure of data processing and show directly in Fig. \ref{exp-pulse2}(b) the results of $\langle i_{HD}^{2}(t) \rangle$ obtained by opening up two arms (blue curve) or blocking one arm (pink curve), respectively. Comparing to Fig.  \ref{exp-pulse2LO}(c), we find the visibility of interference fringe in Fig. \ref{exp-pulse2}(b) is decreased to about 58$\%$ because delay $\Delta T$ is increased from 7.5 ns to 20 ns. When the HD is switched to detectors with response time fast enough to resolve single pulses, no interference can be directly obtained from $\langle i^{2}_{HD}(t) \rangle$, as shown in the upper trace in Fig. \ref{exp-pulse2}(c). However, we can recover the interference through amplitude addition by constructing $i_{+}=i(t)+i(t+\Delta T_e)$ with delay $\Delta T_e \sim \Delta T$. The construction of $i_{+}$ is realized by processing the data of DOS, which is equivalent to split the output of HD with an electronic splitter circuit (ESC) and recombine them by another ESC after introducing an extra electrical delay $\Delta T_e$ in one path, as illustrated in Fig.\ref{exp-pulse2}(a).  The visibility of interference obtained from envelop of $\langle i_{+}^{2}(t)\rangle $ is about 50$\%$,  as shown as the lower trace in Fig. \ref{exp-pulse2}(c). Note that in this case we are also able to recover the interference by taking a time-average of fast voltage pulses at the cost of encoding broadband information in the unbalanced interferometer, which is similar to Fig. \ref{exp-pulse2LO}(e).

\section{Application to quantum systems}

The discussion above uses classical wave theory for simplicity so it applies to the cases with large photon number. In fact, it is similar to the situations in radio astronomy \cite{syn}. In optical regime, however, since optical photons have much larger energy than the radio frequency energy, we have much less photon numbers at the same intensity. For the case of low photon number, we need to use quantum theory of light and it will involve quantum noise, which plays an important role when photon number is low because vacuum quantum noise contributes to homodyne detection. It is well-known that quantum noise can be reduced with squeezed states or entangled states. We can use them for sensing applications.  We present a detailed quantum treatment and give some examples next.

\subsection{Single-photon interference}

\begin{figure}[t]
\includegraphics[width=7cm]{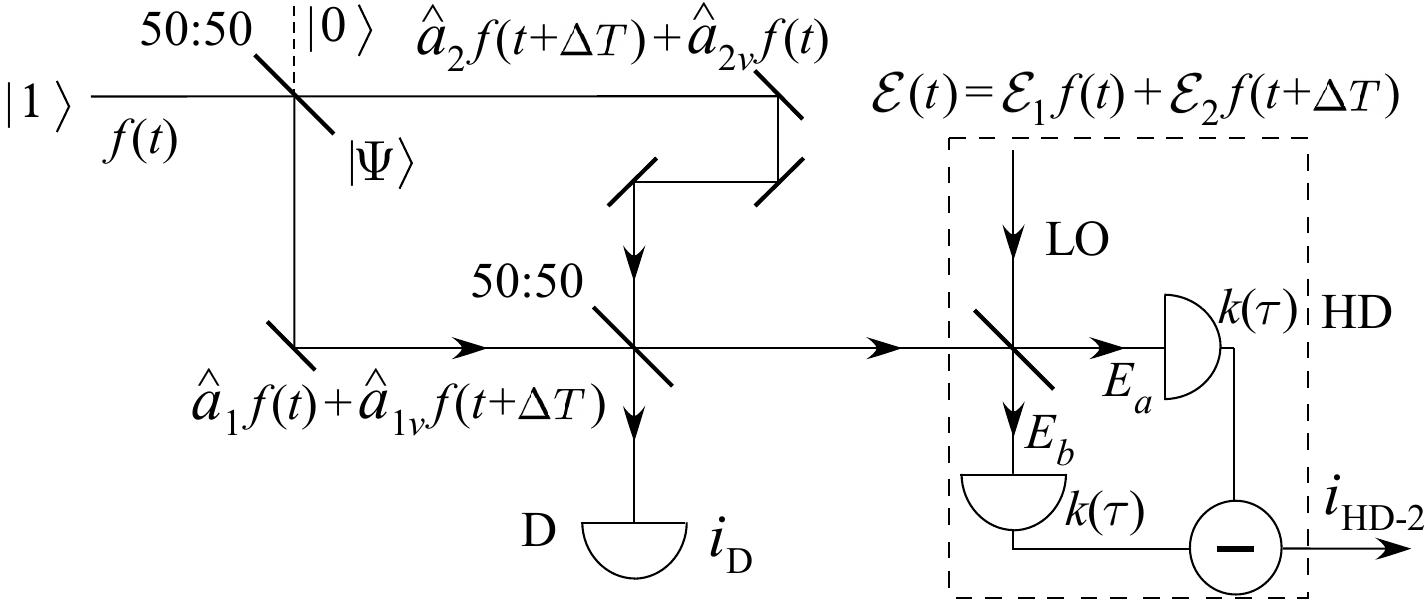}
	\caption{Interference between non-overlapping fields in an entangled single-photon state.}
	\label{SP}
\end{figure}

We first look at a single-photon state, which best describes the particle aspect of light field and usually does not carry any phase information but we send the photon to a 50:50 beam splitter (BS1) and produce an entangled state of single photon: $|\Psi\rangle=(|1\rangle_1|0\rangle_2+e^{i\theta}|0\rangle_1|1\rangle_2)/\sqrt{2}$ (Fig.\ref{SP}). This state was used to demonstrate nonlocality of a single photon \cite{tan}. Now the two fields have coherence and should give rise to interference when recombined by another beam splitter (BS2) as shown in Fig.\ref{SP}. 

To compare with the pulsed case in Fig.\ref{XY}(b), we place the single-photon state in a localized temporal mode with a pulse profile of $f(t)$ (width = $\Delta t$). Here, we assume the spatial mode is well defined and matched so it is not included. Then interference will occur only after proper balance of the paths. Now we add a delay $\Delta T$ to one of the fields so that the pulses do not overlap at the combining beam splitter ($\Delta T \gg \Delta t$). The output field from BS2 is
\begin{eqnarray}\label{Et}
\hat E(t) &=& [\hat a_1f(t)+ \hat a_{1v}f(t-\Delta T)\cr
&&\hskip 0.2in +\hat a_{2v}f(t)+ \hat a_2f(t-\Delta T)]/\sqrt{2},
\end{eqnarray}
where we include the vacuum modes $\hat a_{1v},\hat a_{2v}$ for the corresponding temporal modes in each field.
Then no interference will occur in direct photon counting measurement: $\langle \hat E^{\dag}(t)\hat E(t)\rangle \propto |f(t)|^2+|f(t-\Delta T)|^2$ because of temporal distinguishability: $f^*(t)f(t-\Delta T)=0$. This is so even if we consider the response of the detector: $i_D =\int d\tau k(t-\tau)\langle \hat E^{\dag}(\tau)\hat E(\tau)\rangle \approx k(t) + k(t+\Delta T)$ for urtra-short pulses. We show next how to recover interference by direct amplitude addition via homodyne detection (HD).

To measure the amplitudes of the two temporally separated fields with HD, we use two local oscillator pulses that are matched to the two entangled fields as given in Eq.(\ref{ELO-p}).  Then the vacuum modes will contribute vacuum noise to homodyne detection of two local oscillators.

With response function of $k(t)$ for identical detectors in balanced HD and $\hat E_a(t) =({\cal E} + \hat E)/\sqrt{2}, \hat E_b(t) =({\cal E} - \hat E)/\sqrt{2}$, we can find the output of balanced homodyne detection is \cite{ou-kim}
\begin{eqnarray}\label{HD3}
\hat i_{\rm HD-2}(t) &\propto & \int d\tau k(t+\tau)[\hat E_a^{\dag}(\tau)\hat E_a(\tau)-\hat E_b^{\dag}(\tau)\hat E_b(\tau)]\cr &=&\int d\tau k(t+\tau)\Big\{|f(\tau)|^2|{\cal E}_1| [\hat X_1(\varphi_1)+\hat X_{2v}(\varphi_1)]\cr
 &&\hskip 0.1in + |f(\tau-\Delta T)|^2|{\cal E}_2| [\hat X_2(\varphi_2)+\hat X_{1v}(\varphi_2)]\Big\}/2\cr
&\approx &   k(t)|{\cal E}_1|[ \hat X_1(\varphi_1)+\hat X_{2v}(\varphi_1)]/2\cr
 &&\hskip 0.1in + k(t+\Delta T)|{\cal E}_2| [\hat X_2(\varphi_2)+\hat X_{1v}(\varphi_2)]/2,
\end{eqnarray}
where $\hat X_j(\varphi_j)\equiv \hat a_je^{-i\varphi_j}+\hat a_j^{\dag}e^{i\varphi_j} (j=1,2, 1v, 2v)$ is the quadrature-phase amplitude of each field. In the last line of Eq.(\ref{HD3}), we assume that detector's resolution time $T_R\gg \Delta t$, the width of $f(t)$ so that it cannot resolve the temporal profile of the single pulses, and therefore, $k(t)$ varies much slower than $f(t)$ and can be pulled out of the time integral. Equation (\ref{HD3}) shows that the addition of two amplitudes $\hat X_1(\varphi_1),\hat X_2(\varphi_2)$ or interference between them relies on $k(t)$ or the response of the detector. If the detector is slow so that  $T_R \gg\Delta T$ and thus cannot resolve the two pulses, we have $k(t+\Delta T)\approx k(t)$ and Eq.(\ref{HD3}) gives $i_{\rm HD-2} \propto |{\cal E}_1| \hat X_1(\varphi_1)+|{\cal E}_2| \hat X_2(\varphi_2)+...$, that is, the addition of two amplitudes and interference. Otherwise, there is no overlap between $k(t)$ and $k(t+\Delta T)$ or no addition of two amplitudes. This is complementarity principle at play: low detector resolution means  photon indistinguishability from the two pulses and thus interference or vice versa. To see this quantitatively, we can find the visibility of interference, which will show up in the power of photo-current $\langle i^2_{\rm HD-2}(t)\rangle$ with time integration due to long time average:
\begin{eqnarray}\label{HDi2}
&&\int dt \langle \hat i^2_{\rm HD-2}(t)\rangle_{|\Psi\rangle}\cr && \hskip 0.1in \propto \frac{1}{2} \int dt k^2(t) \Big\{|{\cal E}_1|^2 [\langle\hat X_1^2(\varphi_1)\rangle_{|\Psi\rangle}+1] \cr
&&\hskip 1.4in +|{\cal E}_2|^2 [\langle\hat X_2^2(\varphi_2)\rangle_{|\Psi\rangle}+1]\Big\}\cr
 &&\hskip 0.4in + \int dt k(t)k(t+\Delta T)|{\cal E}_1||{\cal E}_2| \langle\hat X_1(\varphi_1)
  \hat X_2(\varphi_2)\rangle_{|\Psi\rangle}\cr
&& \hskip 0.1in = (3Q_2/2) (|{\cal E}_1|^2+|{\cal E}_2|^2)[1+{\cal V}_{sp} \cos(\theta+\varphi_1-\varphi_2)],~~~~
\end{eqnarray}
with $Q_2\equiv \int dt k^2(t)$ and
\begin{eqnarray}\label{V-sp}
{\cal V}_{sp} \equiv \frac{2|{\cal E}_1||{\cal E}_2|\int dt k(t)k(t+\Delta T)}{3(|{\cal E}_1|^2+|{\cal E}_2|^2)\int dt k^2(t)}.
\end{eqnarray}
The dependence of interference visibility on the resolution of $k(t)$ is obvious in Eq.(\ref{V-sp}) and has a similar form as Eq.(\ref{V}).  

Therefore, interference is recovered even for the case when the photon paths are well distinguished in time. To make it more dramatic, we can rotate the polarization of one of the path by 90 degree so that the photon paths are distinguishable in polarization. In this case, we need to rotate the corresponding LO by 90 degree as well, as in the case of Fig.\ref{XY2}. It is straightforward to show that Eq.(\ref{HD3}) still stands and the rest is the same. 

Notice that we use the wave concept of amplitude here even though we deal with path-distinguished single photon, which is a particle. The interference is due to addition/superposition of the amplitudes of the single-photon field. Thus, the phenomenon is a union of wave and particle in quantum theory. 

Notice that the maximum visibility is only 33\% because of the contribution of the vacuum noise in state $|\Psi\rangle$ to homodyne detection as well as the vacuum noise from the corresponding modes coupled in through BS. This is the difference between the quantum treatment here and the classical treatment in Sect. II. 

\subsection{Reduction of interference visibility due to vacuum noise}

Since we use homodyne detection for the observation of interference, vacuum noise always contributes to the photo-currents no matter what the input field is and will become dominating especially at  low input photon number. Since vacuum does not give rise to interference, the consequence will be the reduction of the visibility of interference. We will analyze this quantitatively next using quantum theory on a general input field.

For simplicity without loss of generality, we only consider the quantum treatment of interference between orthogonal fields $\hat E_{x,y}(t)$ as shown in Fig.\ref{XY}(a).  Quantum mechanically, the photo-current operator after  homodyne detection with two LOs for the two orthogonal polarization modes is given by \cite{ou-kim}
\begin{eqnarray}\label{HD2qs}
\hat i_{HD-2} &=& \hat i_{HDx}(t)+ \hat i_{HDy}(t),
\end{eqnarray}
with
\begin{eqnarray}\label{iHD2}
&& \hat i_{HDx,y}(t) = \frac{|{\cal E}_{x,y}|}{\sqrt{2}} [\hat X_{x,y}(t, \varphi_{x,y})+\hat  X_{x_v,y_v}(t, \varphi_{x,y})],~~~~~
\end{eqnarray}
where $\hat  X_j(t,\varphi_j) \equiv
\hat E_j(t) e^{-i\varphi_j+i\omega_0 t}+\hat E_j^{\dag}(t) e^{i\varphi_j-i\omega_0 t}$ and $\hat  X_{j_v}(t,\varphi_j) \equiv \hat E_{j_v}(t) e^{-i\varphi_j+i\omega_0 t}+\hat E_{j_v}^{\dag}(t) e^{i\varphi_j-i\omega_0 t}$  ($j=x,y$) and ${\cal E}_{x,y}\equiv |{\cal E}_{0}|e^{i\varphi_{x,y}}$ are the amplitudes of the two LO fields of orthogonal polarization but same strength. Here, similar to Eq.(\ref{Et}), we need to consider the vacuum modes for the corresponding polarization in the mixing by the beam splitter.

Assume the two polarization-orthogonal fields are from the 50:50 splitting of a broadband cw field given by
\begin{eqnarray}\label{Ehat}
\hat E(t) = \frac{1}{\sqrt{2\pi}}\int d\omega \hat a(\omega) e^{-i\omega t},
\end{eqnarray}
where we describe the field quantum mechanically with operators. The polarization of one of the fields is rotated by 90 degree after splitting.  Then, $\hat E_x(t) = [\hat E(t) + \hat E_0(t)]/\sqrt{2},  \hat E_y(t) = e^{i\theta} [\hat E(t) - \hat E_0(t)]/\sqrt{2}$ with $\hat E_0(t) = \frac{1}{\sqrt{2\pi}}\int d\omega \hat a_0(\omega) e^{-i\omega t}$ as the vacuum field operators for the unused port of the beam splitter for splitting and $e^{i\theta}$ as an extra phase introduced in y-polarization field, which we will let fluctuate to emulate incoherence.  

By using the commutators calculated as follows
\begin{eqnarray}
[\hat  E(t),\hat  E^{\dag}(t)] &=& \frac{1}{2\pi}\int d\omega d\omega' e^{i(\omega' -\omega)t}[\hat a(\omega), \hat a^{\dag}(\omega')]\cr
&=& \frac{1}{2\pi}\int d\omega d\omega' e^{i(\omega' -\omega)t}\delta(\omega-\omega') \cr &=& \frac{1}{2\pi}\int_{\Delta \omega} d\omega = \frac{1}{2\pi}\Delta \omega \equiv \Delta B\cr
[\hat  E_j(t),\hat  E_j^{\dag}(t)] &=&\Delta B ~~~(j=0, x_v, y_v)
\end{eqnarray}
with $\Delta B$ as the frequency bandwidth of the homodyne detection (the bandwidth comes if we include detector's response),  and 
\begin{eqnarray}\label{X2t}
\langle \hat  X_j^2(t, \varphi_j)\rangle &=& 2\langle \hat  E_j^{\dag}(t)\hat  E_j(t)\rangle + \big\langle[\hat  E_j(t),\hat  E_j^{\dag}(t)]\rangle \cr
&=& R + \Delta B ~~~~(j={x,y})\cr 
\langle \hat  X_{x_v,y_v}^2(t, \varphi_{x,y})\rangle &=& \Delta B\cr
\langle \hat  X_{x}(t, \varphi_{x})\hat  X_{y}(t, \varphi_{y})\rangle &=&  \langle \hat  E_{x}^{\dag}(t)\hat  E_{y}(t)\rangle e^{i(\varphi_x-\varphi_y)}+ c.c. \cr
&&\hskip 0.1 in + \langle [\hat  E_{x}(t),\hat  E_{y}^{\dag}(t)]\rangle e^{i(\varphi_x-\varphi_y)}\cr
&=& R|\gamma|\cos(\varphi_x-\varphi_y+\phi_{\gamma}),
\end{eqnarray}
where $R\equiv \langle \hat  E^{\dag}(t)\hat  E(t)\rangle$ is the photon rate of incoming field, and $\gamma\equiv \langle e^{i\theta}\rangle \equiv |\gamma|e^{i\phi_{\gamma}}$ is the coherence function between x and y-polarized fields,
we can find the power of the photo-current as
\begin{eqnarray}\label{HD2qp}
 \langle\hat i^2_{HD-2} \rangle &=&  \langle [\hat i_{HDx}(t)+ \hat i_{HDy}(t)]^2\rangle\cr
&=& {|{\cal E}_0|^2} (R+2\Delta B) \big[1+{\cal V} \cos (\Delta \varphi+\phi_{\gamma})\big],~~
\end{eqnarray}
where
\begin{eqnarray}\label{V-oe}
{\cal V} \equiv |\gamma|N/(N+2)
\end{eqnarray}
is the visibility of interference with $N\equiv R/\Delta B$ as the total photon number per mode. For a cw field, photons inside a wavepacket of the size of coherence time $T_c=1/\Delta B$ can be regarded as in a single mode so $N  = R T_c = R/\Delta B$ is the photon number per mode. $\Delta \varphi\equiv \varphi_x-\varphi_y$. The term $2\Delta B|{\cal E}_0|^2$ is the contribution of shot noise in homodyne detection with two LOs, which is there even with no input ($R$ or $N=0$) but was ignored in earlier classical treatment. In arriving at the above expression, we assume the incoming field has no phase correlation with the LOs.

\begin{figure}[t]
\includegraphics[width=7.5cm]{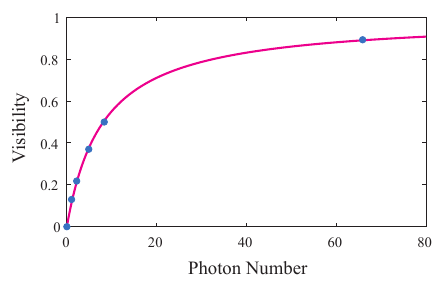}
	\caption{Visibility as a function of average photon number of the input thermal state for the experimental setup in Fig.\ref{exp-pulse2LO}. The solid line is a fit to Eq.(\ref{V-oe}) (see text for details).}
	\label{V-N}
\end{figure}

The visibility reaches the maximum value of $N/(N+2)$
for perfect coherence of $|\gamma| = 1$. The visibility is reduced
due to shot noise contribution in homodyne detection  but  approaches 100\% in the classical limit of large photon number of $N\gg 1$.  Note that the origin of shot noise is the discreteness of photo-electrons and thus exists even for classical fields. The drop of visibility as we reduce photon number is verified experimentally for the scheme shown in Fig.\ref{exp-pulse2LO}. The result is shown in Fig.\ref{V-N} where we plot the visibility of the observed interference fringe as a function of measured average photon number of the input thermal field. The solid curve is a fit to $CN/(CN+2)$ with $C=0.25$, which corresponds to Eq.(\ref{V-oe}) with $|\gamma|=1$ and the losses of the system included.

On the other hand, as is well-known, vacuum noise can be reduced by using quantum states of light. Indeed, squeezed state injection into the unused port can reduce vacuum noise \cite{caves} and quantum entanglement in an SU(1,1) interferometer \cite{ou20}, which uses parametric amplifiers to replace traditional beam splitter as a wave splitter, can have noise reduction in the amplilifier. A recent work has shown that when homodyne detection is used for quantum noise measurement, quantum noise reduction persists even for unbalanced paths in the interferometers \cite{huo22}. Since SU(1,1) interferometers involve more complicated active elements of parametric amplifiers, we will discuss these cases elsewhere.

\section{Relationship to fourth-order interference and interpretation by two-photon interference}

The current technique of amplitude-based interference relies on homodyne detection for amplitude measurement, which involves mixing of interfering fields with strong local oscillator fields (LO). It should be emphasized that interference is not between each of the two interfering fields with its corresponding LO because the phases of the fields and LOs are uncorrelated, and experiments have shown that no interference occurs between the input field and the LO. The interference even appears between two interfering fields in orthogonal modes. Actually, the interference stems from addition/superposition of the amplitudes of the two interfering fields. The role of LOs is to form homodyne detection for the amplitude measurement of each field. 

The view above is of course in terms of wave picture since amplitude concept is only associated with waves. In terms of photon picture, on the other hand, we may associate the phenomena discussed in this paper with fourth-order interference of two photons.
First, notice that there are totally four fields involved. Secondly, the photo-current is an addition of amplitude and thus has a zero average although its envelop shows phase dependent behavior (Fig.\ref{exp-cl}(c), Fig.\ref{unb-cw2}(a), Fig.\ref{unb-i+}(a), Fig.\ref{exp-pulse2LO}(b)).   The interference fringes are more clearly displayed in the average of the power of the current (Fig.\ref{exp-cl}(d), Fig.\ref{unb-cw2}(b), Fig.\ref{unb-i+}(b), Fig.\ref{exp-pulse2LO}(c)). The power is the square of the current, which gives rise to the interference terms in Eq.(\ref{HD2-i}) such as $|{\cal E}_1||{\cal E}_2| \langle\hat X_1(\varphi_1)\hat X_2(\varphi_2)\rangle = \langle {\cal E}_1{\cal E}_2^* \hat E_1 \hat E_2^{\dag}\rangle + \langle {\cal E}_1^*{\cal E}_2^* \hat E_1 \hat E_2\rangle+c.c.$. This involves amplitudes of all four fields. Hence, this is a fourth-order quantity and the phenomenon described here can be associated with fourth-order interference, which is an interference phenomenon involving two photons \cite{ou07}.

With this association, the recovery of interference effect between fields in well-distinguished modes such as orthogonal polarization can be better understood in terms of fourth-order interference of two photons.  As a matter of fact, we encountered something similar in the phenomena of fourth-order interference beyond coherence length \cite{njp,kim,ou22} where interference still occurs even though some of the interfering paths are well separated beyond coherence length. This is because the fourth-order interference term is a fourth-order quantity of the form $\langle V_1(t_1)V_1^*(t_1') V_2(t_2)V_2^*(t_2')\rangle$, which needs no coherence between fields $V_1$ and $V_2$ (Note here that $V_1$ corresponds to LO fields ${\cal E}_1, {\cal E}_2$ and $V_2$ corresponds to $\hat E_1, \hat E_2$). They can thus be fields outside coherence length and there is no need for coherence between the input fields and the LO fields. Here, the fourth-order interference is then a two-photon interference effect with one photon from $V_1$ and other from $V_2$.  So, the interference phenomena discussed in this paper can be regarded as two-photon interference with one photon from one of the two interfering fields and the other from either of the local oscillator fields. This view stands even though the LO fields are strong and contain many photons, in a similar way as Young's double slit interference where photon interferes with itself no matter how strong or weak the interfering fields are.  Furthermore, since two photons are involved, they can be spatially separated so that there are multiple ways to arrange the paths of the two photons \cite{ou22} and it is possible to arrange them in well-distinguished and spatially separated modes for $V_1(t_1)$, $V_1^*(t_1')$ and $V_2(t_2)$, $V_2^*(t_2')$ \cite{franson}.

Traditionally, fourth-order interference involves two detectors and the outcome is the correlation or coincidence between the two intensities measured at the two detectors, which results in the product of the intensities hence also gives fourth-order quantity of field amplitudes.  As described above, the output of HD is in essence a fourth-order quantity of amplitudes but involves only one detector (although balanced HD uses two detectors, it can be done with one detector). While correlation measurement gives direct product of the four amplitudes, HD's amplitude addition, after taking square, can also lead to the product of four amplitudes. The difference lies in one detector in HD versus two detectors in coincidence measurement. Another difference is that traditional fourth-order interference is related to the fourth-order in amplitudes of the interfering fields whereas for homodyne detection, it is still second-order in amplitudes of the interfering fields; the other two amplitudes are from the local oscillator fields. Hence, the detected interference signal is proportional to the square of the photon rate in coincidence measurement but only to the photon rate in homodyne detection.  Under certain circumstance, this can have an advantage in signal level in the latter case when photon rate is low but is still above shot noise level. For example, the total signal level can be ``amplified" by increasing the strength of the local oscillator fields. 

In the interpretation of two-photon interference above, we treat LO fields separately from detectors so Bohr's complementarity principle works when we consider two-photon path indistinguishability at the detectors, which rely on intensity as the observable quantity and still emphasize the particle nature of optical fields.  On the other hand, if we consider the LO fields as a part of the amplitude measurement process, there seems to be a direct challenge to Bohr's complementarity principle: interference is observed even though there is distinguishability by orthogonal polarizations and arrival times of the pulses before measurement. Here, things become a bit complicated because of the involvement of the extra LO fields in homodyne. This is all due to the fact that the fields in optical regime oscillate so fast that no device from current technology can respond, and we have to resort to
homodyne detection for amplitude measurement. Suppose there exists some measurement method that can directly measure electric or magnetic fields without resorting to homodyne detection, then our argument above about complementarity principle applies equally well. This is indeed what happens for radio waves.

The above can be better understood from the point of view of measurement.   Intensity measurement, relying on photo-electric effect, emphasizes the particle nature of light field. It totally loses the wave property such as phase in the process and therefore requires optical field amplitude addition {\it before detection}, leading to the complementarity principle for the detected photon to reveal interference. This can be extended to two-photon interference effect involving coincidence measurement between two detectors. This type of interference effects can be called intensity-based interference. Homodyne detection or any other amplitude measurement method, on the other hand, measures the amplitudes of the field, revealing directly the wave aspect of the field because amplitude is only associated with wave, and therefore does not need optical field addition before detection, which can be achieved {\it after detection}. So, there is no need for complementarity principle in this type of amplitude-based interference phenomena. This is totally different from the traditional intensity-based interference phenomena \cite{wolf} and perhaps is why we can observe interference effects that are not allowed by traditional coherence theory.

\section{Discussion}

It is interesting to note that the visibility of interference depends on the response function $k(\tau)$ of the detectors in both cw and pulsed cases, as shown in Eqs.(\ref{V-p},\ref{V},\ref{qcw-i2pr},\ref{V-sp}). This is confirmed experimentally as shown in Fig.\ref{exp-pulse2LO}(e) for pulsed case and in Fig.\ref{unb-cw2}(c) for the cw case in the unbalanced Mach-Zehnder interferometer. Significantly non-zero visibility is achieved only when there is a substantial overlap between $k(\tau)$ and its delayed part $k(\tau+\Delta T)$ or slow response time, which guarantees the addition of the photo-currents from the fields of the two distinguished arms of unbalanced interferometer for amplitude addition. 
This may be viewed as the extention of Bohr's complementarity principle to photo-current, that is, temporal indistinguishability in photo-currents gives rise to interference. This view encounters another issue in quantum interference:  Dirac statement for photon self-interference. In the single-photon interference discussed in Sect. V, single-photon field is converted to electric currents before superposition to reveal interference. How can classical electric currents preserve quantum superposition? The resolution of this dilemma relies again on our explanation via amplitude-based interference.

The slow response of detectors in the recovery of interference can be equivalent to narrow band filtering of the photo-current, as we discussed in Sect.III. We can actually exploit this ponit of view by doing spectral analysis of the photo-currents from HD measurement of the unbalanced interferometer. Indeed, we can obtain more information about the coherence of the interfering fields. Bcause of the difference and complexity in analysis, we will cover this elsewhere. It turns out that the results are similar to the optical interference effects in frequency domain \cite{wolf,agar,mandel93,rauch}. This is not a surprise since the photon-currents from homodyne detection carry the complete information (both amplitude and phase) of the detected field so spectral analysis of the photo-currents is equivalent to the spectral analysis of the optical field.

It should be noted that the technique of amplitude addition by homodyne detection has been widely used in radio frequency regime. We simply apply it to optical regime. However, because of the huge difference in photon energy between the two spectral ranges, we deal with much less photon numbers in optical regime so that quantum noise becomes a big issue. On the other hand, we can solve this by using quantum states of light for the reduction of quantum noise and still take the advantage of amplitude measurement in the unbalanced interferometer. Since the signal processing technique is quite mature in radio frequency, we can transfer them to optical regime once the quantum noise issue is solved.

It should also be pointed out that homodyne detection (HD) was used in coherent optical communication \cite{tay,gfli,cohcom}, which, however, requires coherence between the signal and local oscillator whereas no coherence is needed here. The HD technique was used here mostly for signal enhancement by the strong local oscillator to overcome electronic noise and dark current noise.

\vskip 0.1in
\noindent {\bf Acknowledgement}

This work was supported in part by the National Natural Science Foundation of China (Grant Nos. 12004279 and 12074283) and by City University of Hong Kong (Project No.9610522), the General Research Fund from Hong Kong Research Grants Council (Nos.11315822, 11301624), and Joint NSFC/RGC Collaborative Research Scheme (12461160325/CRS-CityU103/24).

\vfil
\break

\noindent {\bf Appendix A ~~~Homodyne Detection with Multiple Orthogonal Local Oscillators}

\vskip 0.2 in

Let us first assume the local oscillator field is a multi-mode field described by mode expansion of
\begin{eqnarray}\label{LO-m}
{\cal E}(x) = \sum_m {\cal E}_m u_m(x),
\end{eqnarray}
where $x\equiv \vec x, t$ denotes the space and time coordinates and $\{u_m(x)\}$ consists a set of orthonormal mode functions with $\int  u^*_m(x)u_n(x) dx= \delta_{mn}$. These modes can be polarization modes for vector fields, spatial modes from optical cavity and wave guides, or temporal modes for pulsed fields.

For the field to be detected, we also decompose it into the same set of modes:
\begin{eqnarray}\label{E-m}
\hat E(x) = \sum_m {\hat E}_m u_m(x)
\end{eqnarray}

For balanced homodyne detection, we have the fields after the 50:50 beamsplitter as
\begin{eqnarray}\label{ELO-m}
\hat E_a(x) &=& \sum_m ({\hat E}_m + {\cal E}_m)u_m(x)/\sqrt{2}\cr
\hat E_b(x) &=& \sum_m ({\hat E}_m - {\cal E}_m)u_m(x)/\sqrt{2}.
\end{eqnarray}
The output photo-current from balanced homodyne detection is proportional to
\begin{eqnarray}\label{HD-m}
i_{HD-M} & \propto & \int dx [I_a(x)-I_b(x)] =\int dx [ \langle \hat E_a^{\dag}\hat E_a\rangle - \langle \hat E_b^{\dag}\hat E_b\rangle]\cr
&=&\frac{1}{2} \int dx \sum_{m,n}[ ({\hat E}_m^{\dag}+ {\cal E}_m^*)u_m^*(x)({\hat E}_n+ {\cal E}_n)u_n(x)\cr
&&\hskip 0.4in -({\hat E}_m^{\dag}- {\cal E}_m^*)u_m^*(x)({\hat E}_n- {\cal E}_n)u_n(x)]\cr
&= &\sum_{m,n} ({\cal E}_m^*{\hat E}_n+{\cal E}_n{\hat E}_m^{\dag})\int u^*_m(x)u_n(x) dx \cr
&= &\sum_{m}({\cal E}_m^*{\hat E}_m+{\cal E}_m{\hat E}_m^{\dag})\cr
&=& \sum_{m}|{\cal E}_m|\hat X_m(\varphi_m),
\end{eqnarray}
where we used the orthonormal relation $\int u^*_m(x)u_n(x) dx$ $=\delta_{mn}$ and ${\cal E}_m=|{\cal E}_m|e^{j\varphi_m}, \hat X_m(\varphi_m)\equiv  {\hat E}_me^{-j\varphi_m}+{\hat E}_m^{\dag}e^{j\varphi_m}$.
So, the output current is proportional to the addition of quadrature-phase amplitudes of the orthogonal fields.
Interference should occur between all the orthogonal fields.

\vskip 0.2 in

\noindent {\bf Appendix B ~~~Unbalanced Mach-Zehnder interferometer with quasi-CW  input of fast pulsed train}

\vskip 0.2 in

For quasi-CW field of a pulse train as the input to the unbalanced interferometer in Fig.\ref{unMZ},  the field is described as
\begin{eqnarray}\label{qcw}
E(t) = \sum_j A_j f(t-jT_p)e^{-i\omega_0  (t-jT_p)}
\end{eqnarray}
with $f(t)$ as the normalized pulse profile ($\int dt f^2(t)=1$) and $T_p$ as the separation between adjacent pulses. The width of the pulse ($f(t)$) is typically much shorter than the pulse separation $T_p$.

Let us first consider the case of short coherence time for the source by assuming $A_j$ are independent random variables so that $\langle A_jA_k^*\rangle = I_j\delta_{jk}$, which corresponds to the case of $T_c\ll T_R$ in cw case. Let the unbalance $\Delta T$ of the interferometer be exactly the whole multiple of pulse separation $T_p$: $\Delta T= nT_p$ so that the pulses are still fully overlapped at the second beam splitter (BS2). The delayed field can be written as
\begin{eqnarray}\label{qcw-d}
E(t-nT_p) &=& \sum_j A_j f(t-nT_p-jT_p)e^{-i\omega_0  (t-nT_p-jT_p)+i\theta}\cr &=& e^{i\theta}\sum_j A_{j-n} f(t-jT_p)e^{-i\omega_0  (t-jT_p)}
\end{eqnarray}
where we shift the index of sum by $n$ and introduce an extra phase change $\theta$ in the delayed field (Fig.\ref{unMZ}). With a response function $k(\tau)$ for the detector, the direct detection of the output of the interferometer  then gives the photo-current as
\begin{eqnarray}\label{qcw-iD}
\langle i_{D}(t)\rangle & \propto &\int d\tau k(t-\tau) \langle |E(t)+E(t-nT_p)|^2\rangle/2.~~~~
\end{eqnarray}
Because of the narrowness of $f(t)$ and the independence of $A_j$ ($\langle A_jA_k^*\rangle = I_j\delta_{jk}$), the direct detection current can be calculated as
\begin{eqnarray}\label{qcw-iD2}
\langle i_{D}(t)\rangle & \propto & \int d\tau k(t-\tau)
\sum_j f^2(\tau-jT_p) I_j(1+\delta_{n,0}\cos\theta),\cr
&&
\end{eqnarray}
which shows no interference by direct detection at detector D unless $n = 0$ (balanced case) irrespective of the response of the detector, even though there is superposition of pulses. Now consider homodyne detection with LO as ${\cal E}(t) = {\cal E}\sum_j f(t-jT_p)e^{-i\omega_0 (t-jT_p)}$ with matched pulse profile as the input and fixed amplitude: ${\cal E}=|{\cal E}|e^{i\varphi}$. The photo-current output of the homodyne detection at the output of the interferometer is
\begin{eqnarray}\label{qcw-i}
i_{HDqcw}(t) & \propto &|{\cal E}|\int d\tau k(t-\tau)\sum_j f^2(\tau-jT_p)\cr
&&\hskip 0.5in \times [X_j(\varphi)+X_{j-n}(\varphi-\theta)].~~~~
\end{eqnarray}
Normally, the detector is slow and cannot resolve the profile of the pulses so we can treat $f^2(\tau)$ as a $\delta$-function with normalization $\int d\tau f^2(\tau)=1$:
\begin{eqnarray}\label{qcw-i2}
i_{HDqcw}(t) & \approx &
|{\cal E}|\sum_j k(t-jT_p) [X_j(\varphi)+X_{j-n}(\varphi-\theta)]\cr
&=& |{\cal E}|\sum_j [k(t-jT_p) X_j(\varphi)\cr
&&\hskip 0.3in + k(t-nT_p-jT_p)X_{j}(\varphi-\theta)].
\end{eqnarray}
We made an index change in the last line above. To see interference, we measure the average current power $\langle i^2_{HD-q}\rangle$. Since $\langle A_jA^*_k\rangle =I_j\delta_{jk}$, we have
\begin{eqnarray}\label{qcw-i2p}
\langle i^2_{HDqcw}\rangle & \approx & |{\cal E}|^2\sum_j \langle [k(t-jT_p) X_j(\varphi)\cr
&&\hskip 0.3in + k(t-nT_p-jT_p)X_{j}(\varphi-\theta)]^2\rangle\cr
&\propto & |{\cal E}|^2\sum_j I_j \big[k^2(t-jT_p)+ k^2(t-nT_p-jT_p)\cr
&& \hskip 0.1in + 2k(t-jT_p) k(t-nT_p-jT_p)\cos\theta\big].
\end{eqnarray}
For long time measurement, we integrate over many pulses and arrive at
\begin{eqnarray}\label{qcw-i2p}
\int dt \langle i^2_{HDqcw}\rangle \propto
Q_2 |{\cal E}|^2 \langle I_j\rangle [1+ {\cal V}(nT_p)\cos\theta].
\end{eqnarray}
where $Q_2 \equiv \int dt k^2(t)$. This shows interference with
\begin{eqnarray}\label{Vqcw}
{\cal V}(\tau) \equiv \frac{\int dt k(t) k(t-\tau)}{\int dt k^2(t)}
\end{eqnarray}
as the visibility. From the formula above, we find that interference disappears or ${\cal V}(nT_p)=0$ if $\int dt k(t) k(t-nT_p)=0$ or $k(t)$ and $k(t-nT_p)$ do not overlap. This is the case when the detector can tell the difference between the two paths of interferometer.  On the other hand, we can recover interference by constructing $i_+ = i_{HDqcw}(t) + i_{HDqcw}(t+\Delta T_e)$ with an electronic delay $\Delta T_e$ to compensate the optical delay: $\Delta T_e\sim nT_p$. It is straightforward to show that
\begin{eqnarray}\label{i+sq}
\int dt \langle i^2_+\rangle \propto
Q_2 |{\cal E}|^2 \langle I_j\rangle \Big[1+ \frac{1}{2}{\cal V}(nT_p-\Delta T_e)\cos\theta\Big].~~~~
\end{eqnarray}
This is similar to Eq.(\ref{cw-ipm-sq}) in the cw case.

Next, let us consider the case of long coherence time with $T_c\gg T_R$, which usually occurs in remote sensing but with applicable delay $\Delta T$ limited by coherence time $T_c$. In this case, we can assume $\langle A_jA_{j+q}^*\rangle = I_q \ne 0$ for $q=-M,...,-1,0,1,2,...,M$ with $M=[T_c/T_p]$ but $\langle A_jA_{j+q}^*\rangle = 0$ for $|q|=M+1, M+2, ...$. On the other hand, for large delay with $\Delta T= nT_p \gg T_c$ or $n \gg M$, we find no interference is present in direct detection because of $\langle A_jA_{j+q}^*\rangle = 0$ for $|q|=M+1, M+2, ...$. However, as we will see, interference can recover in homodyne detection with some manipulation.

The output photo-current from HD is still given by Eq.(\ref{qcw-i2}). But for current power $\int dt \langle i^2_{HDqcw}\rangle$, because of the large delay ($\Delta T\gg T_c\gg T_R$), there is no overlap between $k(t-jT_p)$ and $k(t-nT_p-j'T_p)$ even for $|j-j'|\le M$ (for which $\langle X_j(\varphi)X_{j'}(\varphi-\theta)\rangle \ne 0$) so the cross term of interference
\begin{eqnarray}\label{qcw-x}
&&\int dt \sum_j k(t-j T_p) X_j(\varphi)\cr
&&\hskip 0.3in \times \sum_{j'} k(t-nT_p-j' T_p)X_{j'}(\varphi-\theta) =0 .
\end{eqnarray}
This still gives no interference. On the other hand,  we consider the current addition $i_+ = i_{HDqcw}(t) + i_{HDqcw}(t+\Delta T_e)$ with an electronic delay $\Delta T_e$ to compensate the optical delay: $\Delta T_e\sim \Delta T=nT_p$, as before. It has the form of
\begin{eqnarray}\label{qcw-i+}
i_+(t) &\propto &
|{\cal E}| \sum_j [k(t-jT_p) X_j(\varphi)\cr
&&\hskip 0.3in + k(t-nT_p-jT_p)X_{j}(\varphi-\theta)\cr
&&\hskip 0.4in + k(t-jT_p+\Delta T_e) X_j(\varphi)\cr
&&\hskip 0.4in + k(t-jT_p-nT_p+\Delta T_e)X_{j}(\varphi-\theta)].~~~~~~
\end{eqnarray}
When evaluating $\langle i_+^2\rangle$, there will be non-zero cross term between the first and last terms if $\Delta T_e\sim \Delta T=nT_p$:
\begin{eqnarray}\label{qcw-x-delay}
&&\int dt \sum_j k(t-j T_p) \langle  X_j(\varphi)\cr
&&\hskip 0.3in \times \sum_{j'} k(t-j' T_p-nT_p+\Delta T_e)X_{j'}(\varphi-\theta)\rangle \cr
&&= 2\cos \theta \sum_{q=-M}^M \int dt k(t)k(t-q T_p-nT_p+\Delta T_e)2 I_q~~~~~~
\end{eqnarray}
while the other two terms have zero cross terms and give
\begin{eqnarray}\label{qcw-nx}
&&\int dt \langle\Big[\sum_j k(t-j T_p-nT_p) X_j(\varphi-\theta)\Big]^2\rangle\cr
&&= \int dt\langle\Big[ \sum_j k(t-j T_p+\Delta T_e) X_j(\varphi)\Big]^2\rangle\cr
&&=\int dt \sum_j k(t-j T_p+\Delta T_e) \langle X_j(\varphi)\cr
&&\hskip 0.5in \times \sum_{j'} k(t-j' T_p+\Delta T_e)X_{j'}(\varphi) \rangle\cr
&&= \sum_{q=-M}^M \int dt k(t)k(t-j T_p)2 I_q
\end{eqnarray}
to add to the baseline formed by
\begin{eqnarray}\label{qcw-intx}
&&\int dt\Big[ \sum_j k(t-j T_p) X_j(\varphi)\Big]^2\cr
&&= \int dt\Big[ \sum_j k(t-j T_p-nT_p+\Delta T_e) X_j(\varphi-\theta)\Big]^2\cr
&&= \sum_{q=-M}^M \int dt k(t)k(t-q T_p)2 I_q .
\end{eqnarray}
So, the interference is recovered in $i_+$:
\begin{eqnarray}\label{qcw-i+s}
\int dt \langle i^2_+\rangle \propto
Q_2'|{\cal E}|^2 \Big[1+ \frac{1}{2}V_{qcw}(\Delta)\cos\theta\Big]~~~~~~
\end{eqnarray}
with $\Delta\equiv \Delta T_e-nT_p, Q_2'= \sum_{q=-M}^M \int dt k(t)k(t-q T_p)I_q $ and visibility
\begin{eqnarray}\label{Vqcw}
V_{qcw}(\Delta)\equiv \Bigg|\frac{\sum_{q=-M}^M \int dt k(t)k(t-q T_p+\Delta)I_q }{\sum_{q=-M}^M \int dt k(t)k(t-q T_p)I_q }\Bigg|.
\end{eqnarray}

For the case of $T_R\gg T_p$, we can evaluate the integrals in $V_{qcw}(\Delta)$ further by converting sum to an integral with $T_p \rightarrow d\tau$ and $\tau=qT_p$:
\begin{eqnarray}\label{Vqcw}
&& \int dt k(t) \sum_{q=-M}^M k(t-q T_p+\Delta)I_q  \cr
&& ~~~~~~= \frac{1}{T_p} \int dt k(t) \sum_{q=-M}^M k(t-q T_p+\Delta)I_q T_p \cr
&& ~~~~~~\approx \frac{1}{T_p} \int dt k(t) \int d\tau k(t-\tau+\Delta)I(\tau)
\end{eqnarray}
where $I(\tau) = I(qT_p) \equiv I_q$ which has a range of $T_c$. Since $T_c \gg T_R$, $I(\tau)$ can be treated as constant and be moved outside the integral over $\tau$: $\int d\tau k(t-\tau+\Delta)I(\tau) \approx I(t+\Delta) \int d\tau k(t-\tau+\Delta) = I(t+\Delta)Q_1$ and similarly, $\int dt k(t)I(t+\Delta)\approx I(\Delta)\int dt k(t)=I(\Delta)Q_1$. Thus, we have $Q_2' \approx I_0 Q_1^2$ and $V_{qcw}(\Delta)=|I_m/I_0|$ with $m= [\Delta/T_p]$. So, Eq.(\ref{qcw-i+s}) becomes
\begin{eqnarray}\label{qcw-i+s2}
\int dt \langle i^2_+\rangle \propto
Q_1^2I_0 |{\cal E}|^2 \Big[1+ \frac{1}{2}(I_m/I_0)\cos\theta\Big]~~~~~~
\end{eqnarray}
With the definition of $\langle A_jA_{j+m}^*\rangle = I_m$, this is similar to the cw case in Eq.(\ref{cw-ipm-sq}).

\end{document}